%% file: 1-single.tex
\documentclass[9pt,twocolumn,twoside]{pnas-new}

\templatetype{pnasresearcharticle}
\usepackage{amsmath,amssymb,mathtools}
\usepackage{times}
\usepackage{graphicx}
\usepackage{bm}

\setboolean{displaywatermark}{false}

\usepackage{fancyhdr}
\fancypagestyle{firststyle}{
 \fancyfoot[R]{\textbf{\thepage}}
 \fancyfoot[L]{}
 \fancyhead[R]{}
 \fancyhead[L]{}
}

\usepackage{setspace}

\usepackage{color}

\usepackage{url}
\usepackage{hyperref}
\makeatletter
\expandafter\def\expandafter\UrlBreaks\expandafter{\UrlBreaks
\do\a\do\b\do\c\do\d\do\e\do\f\do\g\do\h\do\i\do\j%
\do\k\do\l\do\m\do\n\do\o\do\p\do\q\do\r\do\s\do\t%
\do\u\do\v\do\w\do\x\do\y\do\z\do\A\do\B\do\C\do\D%
\do\E\do\F\do\G\do\H\do\I\do\J\do\K\do\L\do\M\do\N%
\do\O\do\P\do\Q\do\R\do\S\do\T\do\U\do\V\do\W\do\X%
\do\Y\do\Z\do\/\do-}
\makeatother

\usepackage{verbatim}

\newcounter{dataset}
\renewcommand{\thedataset}{S\arabic{dataset}}
\newcommand{\dataset}[2]{%
  \refstepcounter{dataset}%
  {\sffamily\small\bfseries\itshape Dataset  \thedataset\space {\normalfont\sffamily\small\bfseries (\nolinkurl{#1})}}#2
}

\title{Origin of power laws and their spatial fractal structure for city-size distributions }

\author[a,b,1]{Tomoya Mori}
\author[c]{Takashi Akamatsu}
\author[d]{Yuki Takayama}
\author[a]{Minoru Osawa}

\affil[a]{Institute of Economic Research,
Kyoto University. Yoshida-Honmachi, Sakyo-Ku, Kyoto, 606-8501 Japan.
Phone: +81-75-753-7121, E-mail: mori@kier.kyoto-u.ac.jp.}
\affil[b]{Research Institute of Economy, Trade and Industry (RIETI), 11th floor, Annex, Ministry of Economy, Trade and Industry (METI) 1-3-1, Kasumigaseki Chiyoda-ku, Tokyo 100-8901, Japan.}
\affil[c]{Graduate School of Information Sciences, Tohoku University, Aramaki-Aoba, Aoba-Ku, Sendai, Miyagi, 980-8579 Japan.}
\affil[d]{Department of Civil and Environmental Engineering, Tokyo Institute of Technology, 2-12-1 M1-20, Ookayama, Meguro, Tokyo 152-8552, Japan.}
\leadauthor{Mori} 

\authorcontributions{Author contributions: Conceptualization: T.M.; Funding acquisition: T.M., T.A., Y.T., M.O.; Investigation: T.M., Y.T., M.O.; Methodology: T.M., T.A., Y.T.;
Writing: T.M., M.O.}
\authordeclaration{The authors declare no conflict of interest.}
\correspondingauthor{\textsuperscript{1}To whom correspondence should be addressed. E-mail: mori@\@kier.kyoto-u.ac.jp}


\begin{abstract} 
\input{0-abst.tex}
\end{abstract}

\dates{This manuscript was compiled on \today}

\usepackage{bibunits}
\defaultbibliographystyle{unsrt}
\defaultbibliography{many_industries.bib}

\begin{document}

\begin{bibunit}

\maketitle

\thispagestyle{firststyle}
\ifthenelse{\boolean{shortarticle}}{\ifthenelse{\boolean{singlecolumn}}{\abscontentformatted}{\abscontent}}{}

\input{0-text-main.tex}

\acknow{\input{0-ack.tex}}

\showacknow{}

\end{bibunit}

\newpage\null\newpage

\begin{bibunit}
\section*{\huge Supplementary Information}
\input{0-text-si.tex}

\end{bibunit}

\end{document}

%% file: 0-abst.tex
City-size distributions follow an approximate power law in various countries despite high volatility in relative city sizes over time.
Our empirical evidence for the United States and Japan indicates that the scaling law stems from a spatial fractal structure owing to the coordination of industrial locations. While the locations of individual industries change considerably over time, there is a persistent pattern in that more localized industries at a given time are found only in larger cities.
The spatial organization of cities exhibits a hierarchical structure in which larger cities are spaced apart to serve as centers for surrounding smaller cities, generating a recursive pattern across different spatial scales. 
In our theoretical replication of the observed regularities, diversity in scale economy among industries induces diversity in their location pattern, which translates into diversity in city size via spatial coordination of industries and population. 
The city-size power law is a generic feature of Monte-Carlo samples of stationary states resulting from the spontaneous spatial fractal structure in the hypothetical economy. 
The identified regularities reveal constraints on feasible urban planning at each regional scale.
The success or failure of place-based policies designed to take advantage of individual cities' characteristics should depend on their spatial relationships with other cities, subject to the nationwide spatial fractal structure.

%% file: 0-text-main.tex
\dropcap{A}n increasing share of economic activity takes place in cities today \cite{Glaeser-Book2011, Nomaler-et-al-PLOS2014, Balland-et-al-NBH2020}. 
With more than half the world's population residing in urban areas \cite{UN2018}, scientific understanding of cities has become crucial because urban policies may fail to achieve their planned objectives without a solid foundation. 
A natural starting point for any theory is to provide an explanation for the observed regularities. We focus on the well-known fact that 
city-size distribution follows an approximate power law in many countries \cite{Gabaix-Ioannides-2004}. 

The popular theoretical explanation for the city-size power law is random growth models 
\cite{Krugman-JJIE1996, Gabaix-1999,Duranton-RSUE2006, Batty-Nature2006, Duranton-AER2007, Rossi-Hansberg-Wright-REStud2007, Cordoba-JUE2008,   Verbavatz-Berthelemy-Nature2020}, generating the power law as the stationary state of city-size growth processes driven by independent random economic shocks. 
As theories for highly complex systems typically do, random growth models abstract some important aspects of city formation for tractability.  
The most serious simplification is that they ignore economic interactions and spatial relationships between cities, as they assume the growth of cities is mutually independent and cities have no geographical addresses. 
The city-size power laws, however, have a nontrivial spatial structure called the (spatial) \textit{common power law} (CPL). 
Cities exhibit a spatial fractal structure in
that geographically contiguous, rather than random, subsets of the city system follow similar power laws to the whole system \cite{Mori-et-al-PNAS2020}. 
Fig.~\ref{fig:us}A--D shows the case of the United States (US), discussed in detail later. 
Behind the CPL, it is postulated that there is a \textit{spatial-grouping property} (SGP) where smaller cities surround a larger city that serves as their central place \cite{Mori-et-al-PNAS2020}. 
Random growth models do not explain this hierarchical spatial organization because they lack a spatial dimension.
This incompatibility calls for an alternative theory.

\begin{figure}[h!]
    \centering
    \includegraphics[width=\hsize]{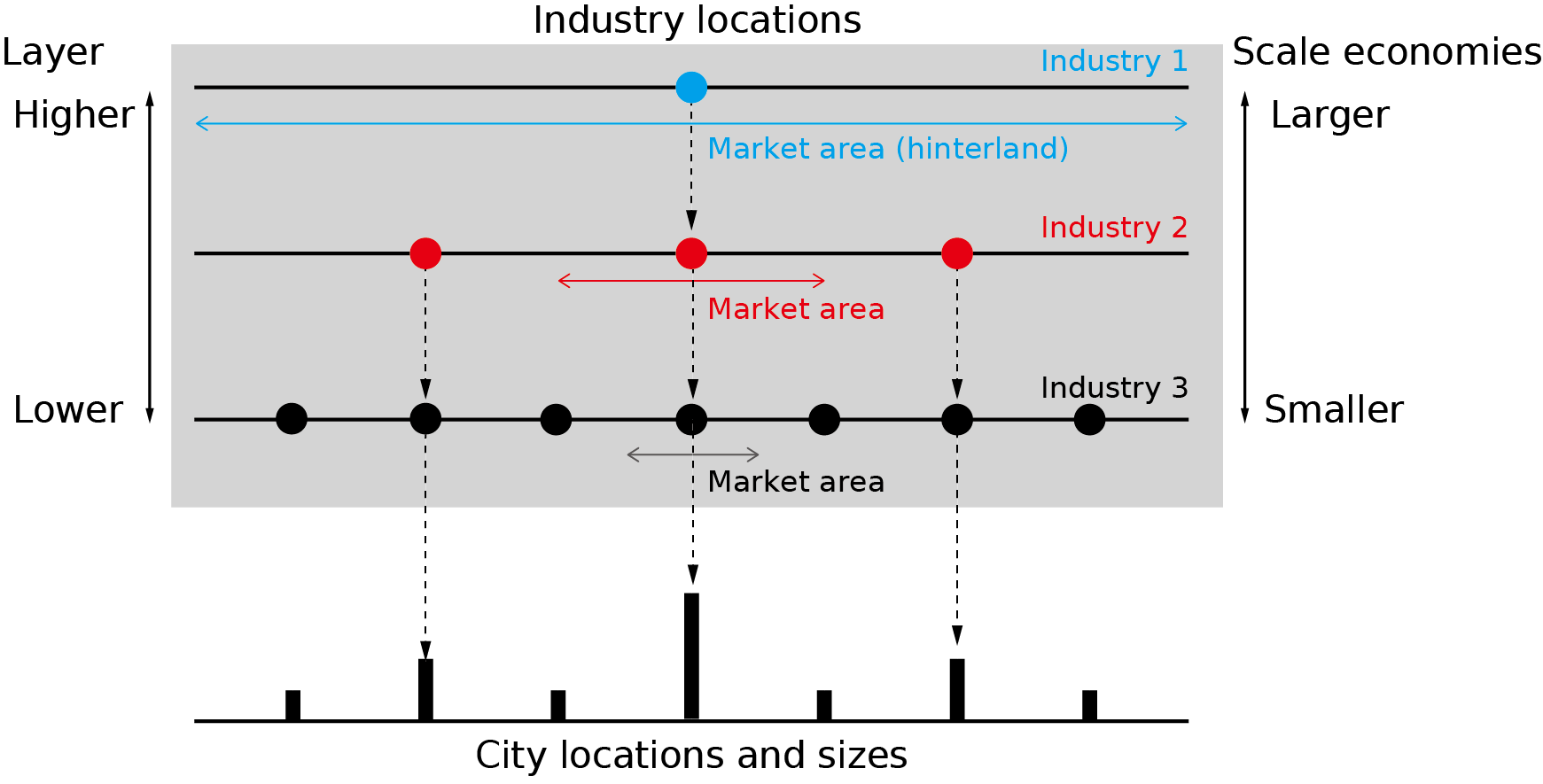}
    \caption{\onehalfspacing 
    Central place theory.}
    \label{fig:CPT}
\end{figure}

Other than random growth models, there are a variety of models 
that explain city-size diversity by  agglomeration economies 
\cite{Duranton-Puga-Book2004} 
combined with innovation \cite{Duranton-AER2007} and comparative advantage \cite{Gaubert-AER2018, Davis-Dingel-JIE2020}.
In relating cities' size and their relative location in space, however, central place theory \cite{Christaller-1933} remains the most prominent conceptual framework \cite{Batty-Longley-1994, Hsu-2012}. 
Central place theory asserts that diverse city sizes accrue from diverse scale economies across industries.
Industries with strong scale economies, such as stock exchanges and musical theaters, exist only in a few cities. They supply their goods and services to those cities'  hinterlands or ``market areas.''
By contrast, industries with weak scale economies, such as bakeries and barbershops, are ubiquitous 
\cite{Mori-et-al-JRS2008,Schiff-JoEG2015,Balland-et-al-NBH2020}. 
Consequently, city systems spontaneously exhibit hierarchical layers in which the smaller city's set of industries is a subset of the larger city's (henceforth, the \textit{hierarchy property}, HP). 
Equivalently, layers are characterized by sets of industries that do not exist in lower layers (Fig.~\ref{fig:CPT}). 
In each layer, every city is associated with its hinterland, where it supplies characteristic goods to smaller cities in lower layers, where those goods are unavailable. 
Consequently, larger cities would be evenly located among smaller cities, displaying a recursive structure, the SGP.
As the SGP and the HP characterize the spatial fractal structure postulated by central place theory, we call their combination the \textit{central place hierarchy} (CPH). 
We argue that the CPH results in diverse city sizes and power laws, as suggested in the literature \cite{Beckmann-1958,Batty-Longley-1994,Hsu-2012}.

In this study, we demonstrate the empirical relevance of central place theory as a mechanism behind the CPL through two contributions. 
First, we offer quantitative evidence for the CPH in the US city system. The results for Japan are reported in Sec.\,\ref{sec:supp-text}\ref{sec:si-jp} in the Supplementary Information (SI).
Specifically, we show that the SGP holds recursively 
in the sense that 
most cities, regardless of size, are surrounded by cities smaller than themselves,  indicating the fractal structure of the whole city system. 
We also show that almost all the secondary and tertiary industries in typical industrial classifications exhibit strong hierarchical coordination as presumed in central place theory, delivering robust evidence for the HP. 

Second, we provide a microeconomic theory to reproduce the empirical regularities. 
We develop a many-industry general equilibrium model of city formation, incorporating mechanisms that produce the CPH in its stationary states.
When the real-world distribution of the scale-economy parameter is imposed, the model qualitatively reproduces the CPL as a generic feature of numerous stationary states.

Unlike random growth models, our model can quantitatively assess economic policies as it considers essential economic interactions such as labor markets and goods trade between locations. 
It is a step toward urban and regional policy evaluations based on counterfactual simulations where empirical regularities (e.g., the CPL) are imposed as indispensable side constraints.

\section*{Common power law for US city sizes}
\begin{figure*}[h!]
\centering{}\includegraphics[width=17.1cm]{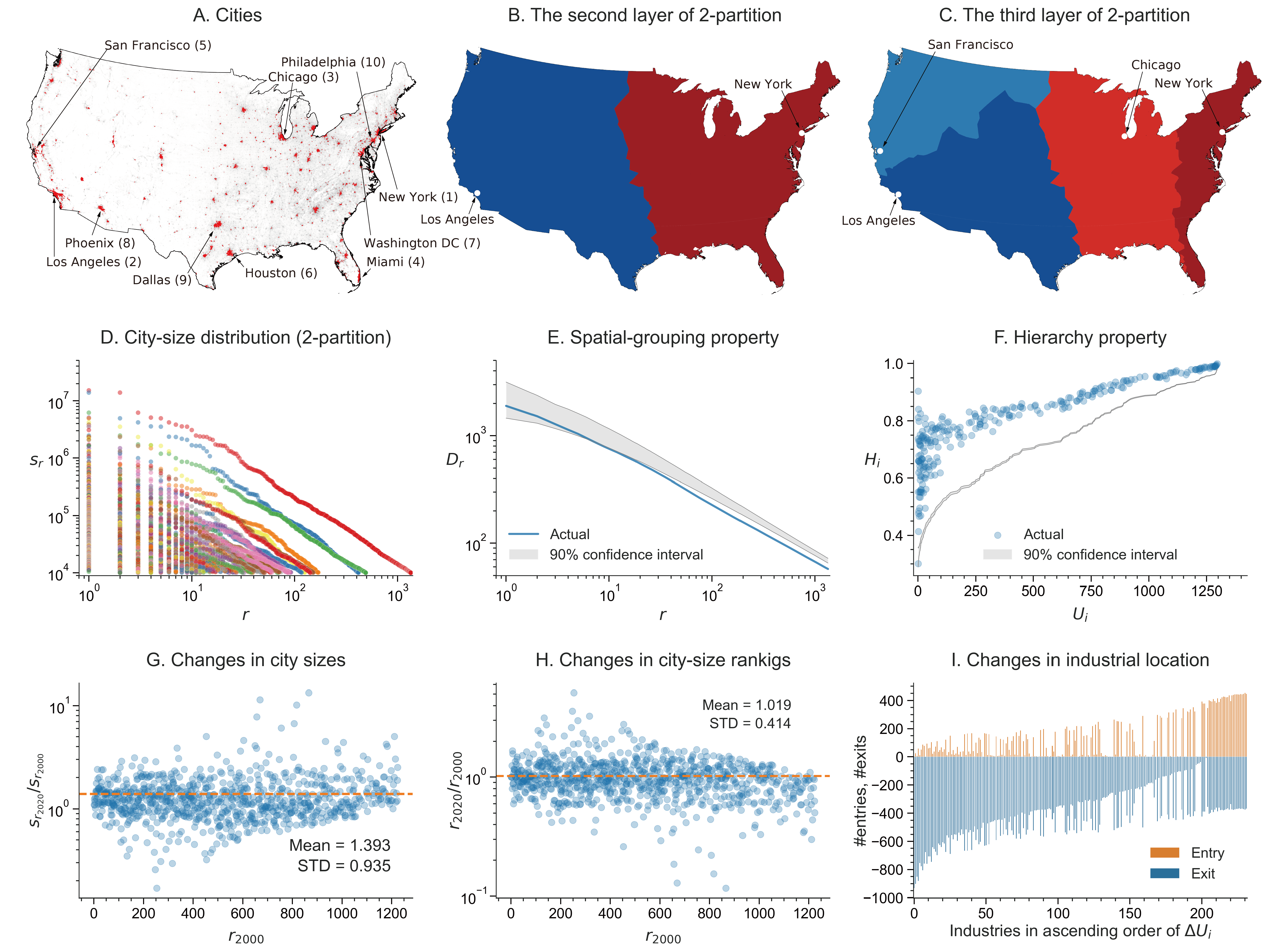}
\caption{(A-F) considers the continental-US cities in 2020. (A) Red areas indicate all the 1,344 cities; darker gray corresponds to a larger population per 1 km-by-1 km grid. 
The 10 largest cities are indicated with their population rankings in parentheses. 
(B,\,C) The second and third layers of the 2-partition of cities.
(D) City-size distributions in the 2-partition cells.
(E) $D_r$ and the 90-percent confidence interval of $\tilde{D}_r$.
(F) $H_i$ against the number $U_i$ of choice cities for the 258 four-digit NAICS industries together with the 90-percent confidence interval of $\tilde{H}_i$.
(G,\,H) City-size and city-size-ranking ratios between 2020 and 2000 against city-size rankings, respectively, in 2000 for the 852 cities that existed in both years.
(I) The numbers of choice cities in 2020 but not in 2000 (entry) and those in 2000 but not in 2020 (exit) of each of the 232 four-digit NAICS industries that existed in both years. Industries are ordered along the horizontal axis in the ascending order of the net change in the number of choice cities between 2000 and 2020.}
\label{fig:us} 
\end{figure*}

To formalize the notion of city-size CPL,
we first revisit Mori et al.~\cite{Mori-et-al-PNAS2020}'s approach to statistically test the CPL using the US data. 
Below, bold capital italic characters (e.g., $\bm  U$) represent a set, whereas regular italic characters (e.g., $U$) represent its cardinality (i.e., $U = |\bm U|$). 
In the empirical analysis, a \emph{city} indicates a contiguous area with a density of at least 1,000 people per square kilometer, yielding a total population of at least 10,000 (Fig.~\ref{fig:us}A).
Let $\bm  U\equiv \{1,2,\ldots, U\}$ be the set of all cities in the country ranked by population size. 
In Fig.~\ref{fig:us}B, the red plots, above all plots in other colors, show the rank--size plot for all continental US cities, demonstrating a clear log-linear relation towards its tail. 

As the relevant subsets of $\bm {U}$ for the CPL test, we consider a collection of sub-city systems consistent with the SGP. 
Following \cite{Mori-et-al-PNAS2020}, we consider an \emph{$L$-partition}, a hierarchical Voronoi partition of $\bm {U}$ with a fixed integer $L \ge 2$. 
We take the $L$ largest cities (central places) and assign all other cities to the closest among them, yielding $L$ disjoint subsets of $\bm U$. 
Recursively generating a new partition for each subset concerning the $L$ largest cities in the subset, we obtain a unique hierarchical partition of $\bm U$. 
Figs.~\ref{fig:us}B and C show the second and third layers of the continental US's 2-partition, respectively.
For each city $j\in \bm U$, $j$-\emph{hinterland} is the highest-layer subset of the hierarchical partition that $j$ serves as the central place. 
For example, New York's hinterland is the set of all US cities, that of Los Angeles is the set of cities in the  blue cell of Fig.~\ref{fig:us}B, and that of San Francisco is the lighter-blue cell of Fig.~\ref{fig:us}C.
The collection of $j$-hinterlands is taken as the relevant city subsets for the CPL test. 
Fig.\,\ref{fig:us}D shows the city-size distributions in the 2-partition.

The CPL coefficient is estimated by a categorical regression with fixed effects for hinterlands. 
To obtain a collection of rank-size data $\{(\ln r_{ij},\ln s_{ij})\}$, $r_{ij}$ denotes the rank of city $i$ in $j$-hinterland and $s_{ij}$ its population size. 
The regression model is given by 
\begin{equation}
    \ln s_{ij} 
        = 
         \beta_{1}
         - \theta\ln(r_{ij} - 0.5)
         + \sum_{k\geq 2}\beta_{j}\delta_{j}(k)
         + \varepsilon_{ij}.
    \label{eq:cat-regression}
\end{equation}
We take central place $1$ (New York) as the reference and $\delta_{j}(k)$ are indicator variables for other central places $k\geq 2$ such that $\delta_{j}(k)=1$ if $k=j$ and zero otherwise. $\varepsilon_{ij}$ is an error term. 
The subtraction of $0.5$ from $r_{ij}$ is a bias correction \cite{Gabaix-Ibragimov-2011}. 
We hypothesize the slope coefficient $\theta$ to be the same for all $j\geq 1$ under the CPL. 
The estimated slope coefficient for the US city system under the 2-partition is $1.078$ in 2020.

The hierarchical and spatially contiguous nature of an $L$-partition plays a significant role in the commonality of the power-law coefficient among the hinterlands.
As the test static for the CPL, we use the fit of the regression model Eq.~\ref{eq:cat-regression} evaluated by the average mean squared error (MSE). 
We apply a permutation test to show that the model fit deteriorates significantly when we perturb the underlying city subsets. 
Perturbed city subsets are constructed by considering counterfactual partitions, in which the Voronoi partition of a city subset concerning its $L$ largest cities is replaced by a random one that ignores the spatial organization of cities. 
Specifically, cities in a given subset are assigned randomly to their $L$ largest cities, while the number of cities in each resulting $L$ lower-layer subset remains unchanged from the original $L$-partition. 
The null hypothesis of the CPL test is that the MSE of the $L$-partition and those of the random partitions belong to the same statistical population. 
The CPL implies that the MSE value under the $L$-partition is significantly smaller than those under random partitions. 
The null hypothesis is rejected by the one-sided test at the 1\% level under $L=2,\ldots,6$ in 2000, 2005, $\ldots$, 2020.
The test results suggest that the observed CPL is unlikely to realize under random growth processes.
See SI for the details and the hypothesis-test framework in this article.

\section*{Central place hierarchy}

For central place theory to be a plausible explanation for the city-size power law, including the CPL, its fundamental premise, the CPH, must be observed in real-world data. 
We show that the hierarchical spatial organization of the US city system is consistent with the CPH by testing the SGP for cities and the HP for the industrial location pattern.

\smallskip
\noindent {\sffamily\bfseries Spatial-grouping property.}
To test the SGP, we introduce a scalar measure for the overall spatial organization of cities. 
As a preparation, we define the average proximity between the $r$ largest cities, $\bm  U_r$, and the rest of the smaller cities, $\bm U \backslash \bm U_r$, by the average of the road distance from each city of the latter to the closest one in the former,
\begin{equation}
    d_r = \frac{1}{U-U_r}\sum_{u\in \bm  U\backslash \bm U_r} \min_{u'\in \bm  U_r} \mathrm{distance}(u,u'),\quad r< U.
    \label{eq:spacing-out}
\end{equation}
To gauge the overall consistency of the $r$ largest cities' location pattern with the SGP by a single statistic, we use the average of $d_r$ for the top $r$ cities: 
\begin{equation}
    D_{r} =\frac{1}{r}\sum_{r' \leq r} d_{r'},\quad r<U.\label{eq:SGP}
\end{equation}

To construct counterfactuals of $\{D_r\}_{r<U}$, we randomly order all cities, and take the set of the first $r=1,\ldots,U-1$ cities, $\tilde{\bm U}_r$ as the counterfactual of $\bm U_r$. 
The correponding counterfactuals $\tilde{d}_r$ of $d_r$ are computed by Eq.~\ref{eq:spacing-out} replacing $\bm  U_r$ with $\tilde{\bm  U}_r \subseteq \bm  U$. 
The SGP implies that $d_r < \tilde{d}_r$ for a wide range of $r$'s simultaneously, because it requires that larger cities are evenly located among smaller cities (Fig.~\ref{fig:CPT}).

Fig.~\ref{fig:us}E shows $D_r$ of the US cities for each $r<U$ in 2020 together with the 90-percent confidence interval of $\tilde{D}_r$, constructed from 1,000 counterfactual samples.
We find that $D_r$ is smaller than the 5-percentile values of $\tilde{D}_{r}$ for all $r\geq 8$, rejecting the null hypothesis at the 5\% significance level that $D_{r}$ and $\tilde{D}_{r}$ belong to the same statistical population.
The largest cities ($r\leq 7$) do not satisfy the SGP because they are typically located at the coasts for better access to the rest of the world, which is outside the scope of classical central place theory that assumes featureless geography. 
The remaining medium-sized cities are, however, domestically oriented and exhibit the SGP with statistical significance. 
Similar results are obtained for the US cities in 2000, 2005, \ldots, 2015 (see Sec.\,\ref{sec:supp-text}\ref{sec:si-us} of SI and Fig.~\ref{fig:sgp-2000_us} for the case of 2000).

As an alternative test for the SGP, we test for \textit{central-place property} (CPP), which asserts that a city in a higher layer in the $L$-partition is larger than those in lower layers (see Sec.\,\ref{sec:supp-text}\ref{sec:si-us} of SI).

\smallskip
\noindent{\sffamily\bfseries Hierarchy property.}
To test the HP, we use the industry location data of the 258 four-digit secondary and tertiary industries in the North American Industrial Classification System (NAICS).  
Let $\bm  I$ be the set of all industries.
If industry $i\in \bm  I$ is present in a city, we call this city a \emph{choice city} of $i$ and denote the set of these cities by $\bm  U_i\subseteq \bm  U$. 
Industry $i$ is more localized than industry $j$ if $U_i<U_j$ and more ubiquitous if $U_i > U_j$. 
Industry $i$'s consistency to the HP can be measured by
\begin{equation}
    H_i = \frac{1}{J_i U_i}\sum_{j \in \bm  J_i } \left| \bm  U_i \cap \bm  U_j\right|
    \label{eq:h}
\end{equation}
where $\bm  J_i \equiv \{j\in \bm  I\setminus \{i\} : U_i \leq U_j\}$ is the set of industries at least as ubiquitous as $i$. 
The maximum value for $H_i$ is 1 when coordination is perfect ($\bm  U_i\subseteq \bm  U_j$ for all $j\in \bm  J_i$).
To gauge the significance of $H_i$-values, we compute $\tilde{H}_i$, counterfactual $H_i$-values, by replacing $\bm U_i$ and $\bm U_j$ in Eq.~\ref{eq:h} with randomly chosen city subsets of the same sizes.
Fig.~\ref{fig:us}F plots $H_i$ for each industry in 2020 together with the 90\% confidence interval of $\tilde{H}_i$, constructed from 1,000 counterfactual samples.
$H_i$ for most industries are significantly higher than their counterfactual values, with an average of 0.795.

In addition, if $\bm I_u$ represents the industry set in city $u\in \bm  U$, the HP implies $\bm I_u\subseteq \bm  I_v$ for $u,v\in \bm  U$ if $I_u \leq I_v$.
The empirical pattern is consistent with the HP, as we observe a positive association between industrial diversity and city size, with a Spearman's rank correlation of 0.794. 
The HP is persistent between 2000 and 2020. 
See Sec.\,\ref{sec:supp-text}\ref{sec:si-us} for the results under alternative industry-aggregation levels in 2020 and 2000 (Fig.~\ref{fig:us_hp_2020} and \ref{fig:us_hp_2000}, respectively).

\section*{Volatility of city sizes and industrial location}
Despite the persistent CPL and HP between 2000 and 2020, relative city sizes fluctuated substantially in the same period (Figs.~\ref{fig:us}G and H). 
Also, the locations of individual industries changed considerably. 
Fig.~\ref{fig:us}I shows the changes in the sets of choice cities of the 232 four-digit NAICS industries that existed in both years. The numbers of entries to cities and exits from cities for industries between 2000 and 2020 are 123 and 425 on average, respectively.
Some industries have exited the urban market entirely, whereas others relocated to different cities.
In particular, the exiting industries in this period reflect the influence of internet. 
For example, ``Book Stores and News Dealers,'' ``Florists,'' ``Travel Arrangement and Reservation Services'' are among the ones that exited from the majority of cities (see Table \ref{tb:industry-churning}).
The volatility of city sizes and industrial location has been reported under different definitions of cities and industrial location  \cite{Batty-Nature2006,Verbavatz-Berthelemy-Nature2020,Dumais-Ellison-Glaeser-REStat2002, Duranton-AER2007, Findeisen-Suedekum-JUE2008}.
Despite these considerable changes, the CPL and CPH remains stable, suggesting that these regularities are more instantaneous than dynamic. This finding motivates us to formulate a static central-place model to explain the observed regularities.

\section*{A model of central places and power law}
The US city system exhibits both the CPL and CPH, suggesting industrial diversity as the key underlying factor. 
For qualitative reproduction of the empirical facts, we propose an economic model incorporating central place theory. 
We construct it using standard model components from the spatial economics literature \cite{Fujita-Krugman-Mori-1999, Tabuchi-Thisse-2011}.
See SI for details of the mathematical model and simulations.

There are $R$ viable locations for cities and $I$ industries whose sets are denoted by $\bm  R$ and $\bm  I$, respectively. 
There are footloose agents that maximize their payoff by choosing the location they reside in and their work industry. 
Let $h_{ir}$ be the number of agents who live in location $r$ and are employed in industry $i$. 
A stationary state of the model, or a \emph{spatial equilibrium}, is an $R \times I$-dimensional distribution $\bm  h = \{\bm  h_i\}_{i\in\bm  I} = \{h_{ir}\}_{i\in \bm  I, r\in\bm  R}$ under which each agent maximizes their payoff. 
Let a \emph{city} be a location $r$ for which $h_{ir} > 0$ for some $i$. 
The set of cities corresponds to $\bm  U$ in our empirical analysis. 
The size of city $r$ under a distribution $\bm  h$ is $\sum_{i\in\bm  I} h_{ir}$. 
 
As for industrial diversity, we focus on the variation in product differentiation among other conceivable factors discussed in the literature \cite{Fujita-Krugman-Mori-1999, Hsu-2012}. 
In each industry, every firm produces a differentiated variety of industry-specific goods. 
Different industries are subject to different levels of \emph{transport freedom} between cities.  
Let $\tau_{rs} > 1$ be the level of physical trade cost between locations $r$ and $s$, common to all industries.   
The transport freedom between $r$ and $s$ for industry $i$ is $\phi_{i,rs} = \tau_{rs}^{1 - \sigma_i} \in (0,1)$, where $\sigma_i > 1$ is the substitution elasticity between product varieties in industry $i$. 
The parameter $\sigma_i$ is interpreted as the degree of differentiation between the different varieties in industry $i$; a higher substitutability 
(larger $\sigma_i$) means less distinguishable varieties.  
If $\sigma_i<\sigma_j$, then $\phi_{i,rs} > \phi_{j,rs}$, reflecting that trade volume between cities decreases faster for an industry supplying more substitutable goods.
A longer transportation distance implies a higher delivered price, and hence, locally produced varieties are more attractive than distant ones if product varieties are less differentiated. 
Consequently, the effective supply area of an $i$-industry firm tends to become smaller when $\sigma_i$ is higher.

With appropriate scaling of variables, each mobile agent corresponds to exactly one firm and $h_{ir}$ also represents the number of firms in industry $i$ located in city $r$.  
Then, $M_{ir}(\bm  h_i) = \sum_{s \in \bm  U} \phi_{i,rs} h_{is}$ is a measure of consumers' \emph{market access} to industry $i$ in location $r$. 
Consumers prefer locations with better access to a wider range of goods, measured by aggregate market access $M_r(\bm  h) = \sum_{i\in \bm  I} \frac{1}{\sigma_i - 1} \log M_{ir}(\bm  h_i)$. 
A large concentration of firms promotes a large concentration of consumers, inducing an agglomeration of firms and consumers to form cities.
As consumers are common to all industries, there is an incentive for firms in different industries to spatially coordinate.

From $i$-industry firms' perspective, $M_{ir}(\bm  h_i)$ is a measure of competition in city $r$. 
For a given population distribution, some cities are not profitable for industry $i$, because they are too close to $i$-industry's competitors in nearby cities. 
In each industry, concentrations of firms must be sufficiently separated geographically to avoid competition. 
The required geographical separation depends on the $\sigma_i$-value (Fig.~\ref{fig:CPT} and \ref{fig:market-area}). 
If $\sigma_i < \sigma_j$, then $\phi_{i,rs} > \phi_{j,rs}$ and thus, for any city $r$, $M_{ir}(\bm  z) \ge M_{jr}(\bm  z)$ for any $\bm  z \ge \bm  0$, meaning that competition between firms tends to be more intense for industries with smaller $\sigma$ (more differentiated goods). 
Firms in small-$\sigma$ industries have incentives to enter fewer locations than those in large-$\sigma$ industries. 
Consequently, if $\sigma_i < \sigma_j$, industry $i$ operates in a smaller number of locations than industry $j$, or choice cities tend to be more spaced apart for industry $i$ than for industry $j$, resulting in the HP.

\section*{Simulations}
In simulations, we consider a hypothetical geography in which $1\rm{,}024$ locations are equidistantly placed along the circumference. For a given assignment of parameter values, there are numerous equilibria and their mathematical characterizations 
are impossible. 
We conduct a Monte Carlo experiment that samples $1\rm{,}000$ equilibrium spatial distributions for each given number of industries ($I = 4,16,64,256$).

The key parameters are $\{\sigma_i\}_{i\in\bm  I}$ that represent industrial diversity in scale economies, the source of the model's city-size diversity. 
The city-size power law can emerge when the recursive structure, or the CPH (Fig.~\ref{fig:CPT}), is formed down to a sufficiently deep hierarchy \cite{Hsu-2012}. 
For this, $\{\sigma_i\}$ must take sufficiently diverse values. 
Although we do not know theoretically plausible distributions for $\{\sigma_i\}$, empirical estimates are available. 
We use the set $\bm {\Sigma}$ of substitution elasticity estimates for 13,930 products imported into the US \cite{Broda-Weinstein-2006}, which have a considerably wide range ($1.03\leq \sigma_i \leq 4\mathrm{,}303$). 
Fig.~\ref{fig:sm}A shows the distribution of the price markup ratio to the marginal cost, $\sigma_i/(\sigma_i-1)$, a measure of industries' scale economies (see also Sec.\,\ref{app:markup} and Fig.\,\ref{fig:markups} in SI). 
These are the only readily available large distributional data that can be considered to proxy the real-world diversity in scale economies, which motivated our formulation. 
To obtain each equilibrium sample, we draw $\{\sigma_i\}_{i\in\bm I}$ randomly from $\bm {\Sigma}$ and then find a locally stable equilibrium from a random initial state.

The CPL and CPH emerge as generic properties of spatial equilibria for almost all 1,000 sample equilibria when $I$ is large, that is, under sufficient diversity in firms' scale economy. 
As a demonstration, we focus on the first equilibrium sample under $I = 256$. 
Fig.~\ref{fig:sm}B displays the spatial distribution of cities, and the second and third layers of the 2-partition for the equilibrium. 
The CPL of the city-size distributions in the hypothetical economy (Fig.~\ref{fig:sm}C) is as clear as that in the US (Fig.~\ref{fig:us}D). 
The results on the CPH shown in Fig.~\ref{fig:sm}D and E correspond to Fig.~\ref{fig:us}E and F for the US case, respectively.
The SGP is significant for all $r\in [4,373]$ (Fig.\,\ref{fig:sm}D).

The deviation from the SGP at small and large $r$ is partly due to the specificity of the hypothetical location space (Fig.\,\ref{fig:sm}D).
For $r=1$ and $2$, the actual and counterfactual values of $D_r$ must be similar because the location space is circular.
Larger inter-city distances among small cities (for $r\geq 374$) are observed because %
the city spacing is overstated for the smallest cities in the theoretical model since it is lower-bounded by the spacing of the given viable locations.

For all $i$, $H_i$ is close to $1$, indicating almost maximal hierarchical coordination among industries in equilibrium (Fig.~\ref{fig:sm}E).
The results are similar for all other samples even though we draw different values for $\{\sigma_i\}$ and the initial spatial distribution for each sample.
The HP is persistent for different $I$s, as the average $H_i$-value among all equilibrium samples is greater than $0.984$ for all $I=4,16,64,256$ (Fig.~\ref{fig:hp-sm}).
However, the HP translates to a large diversity in city size and spacing only when there is a large diversity in scale economies, that is, when $I$ is large (Fig.~\ref{fig:size-diversity} and \ref{fig:app-sgp-sm}).
When $I$ is large, $\sigma_i$ values tend to have a large variation, which in turn translates to a large variation in the number of choice cities among industries (Fig.~\ref{fig:markups}).
Through the HP, a city's size and industrial diversity exhibit a positive correlation, increasing in $I$ (Fig.~\ref{fig:div-sm}).
Their average Spearman's rank correlation across the $1\mathrm{,}000$ equilibrium samples increases from $0.554$ under $I=2$ to $0.942$ under $I=256$. 
The CPH becomes more apparent under a larger $I$. See Sec.\,\ref{sec:supp-text}\ref{sec:si-sm} and Fig.\,\ref{fig:summary-sm} of SI for extensive evidence of the consistency between the hypothetical economy and the CPH.

Consequently, the CPL emerges under sufficiently large $I$ (Fig.~\ref{fig:cplsm}A--C and \ref{fig:sm}B). 
For $I\geq 128$, the null hypothesis of the CPL test is rejected at the 0.05 level for more than 95\% of the equilibrium samples obtained under $L=2,\ldots,6$ (Fig.~\ref{fig:cplsm}D), indicating the CPL being a generic property of stable equilibria in the hypothetical economy.

\section*{Discussions} 

We have empirically and theoretically shown that the city-size power law may have its root in spontaneous economic forces that govern industrial agglomeration as envisaged by central place theory. 
Our proposed model reproduces the CPL and CPH simultaneously, 
without any exogenous differences across locations owing to geographical advantages, available natural resources, and other historical factors. 
Although exogenous differences may determine the actual locations of cities in the real world, system-wide regularities, such as the CPL, may arise from endogenous economic forces.

Our results generalized the results of \cite{Hsu-2012} in two critical aspects.
First, given a sufficient diversity in scale economies, we showed based on the Monte-Carlo sampling  that the CPL and CPH hold in most of the numerous different equilibria realized by the self-organization of firms and populations from their random initial distributions. 
While \cite{Hsu-2012} formally showed the existence of such an equilibrium in a stylized model, the stability and potential multiplicity of equilibria were left unknown.
In our approach, it is possible to evaluate the formation of the CPL and CPH in the empirical and theoretical city systems using the same statistical test, allowing a direct comparison between them. %

\begin{figure*}[h!]
    \centering
    \includegraphics[width=17.1cm]{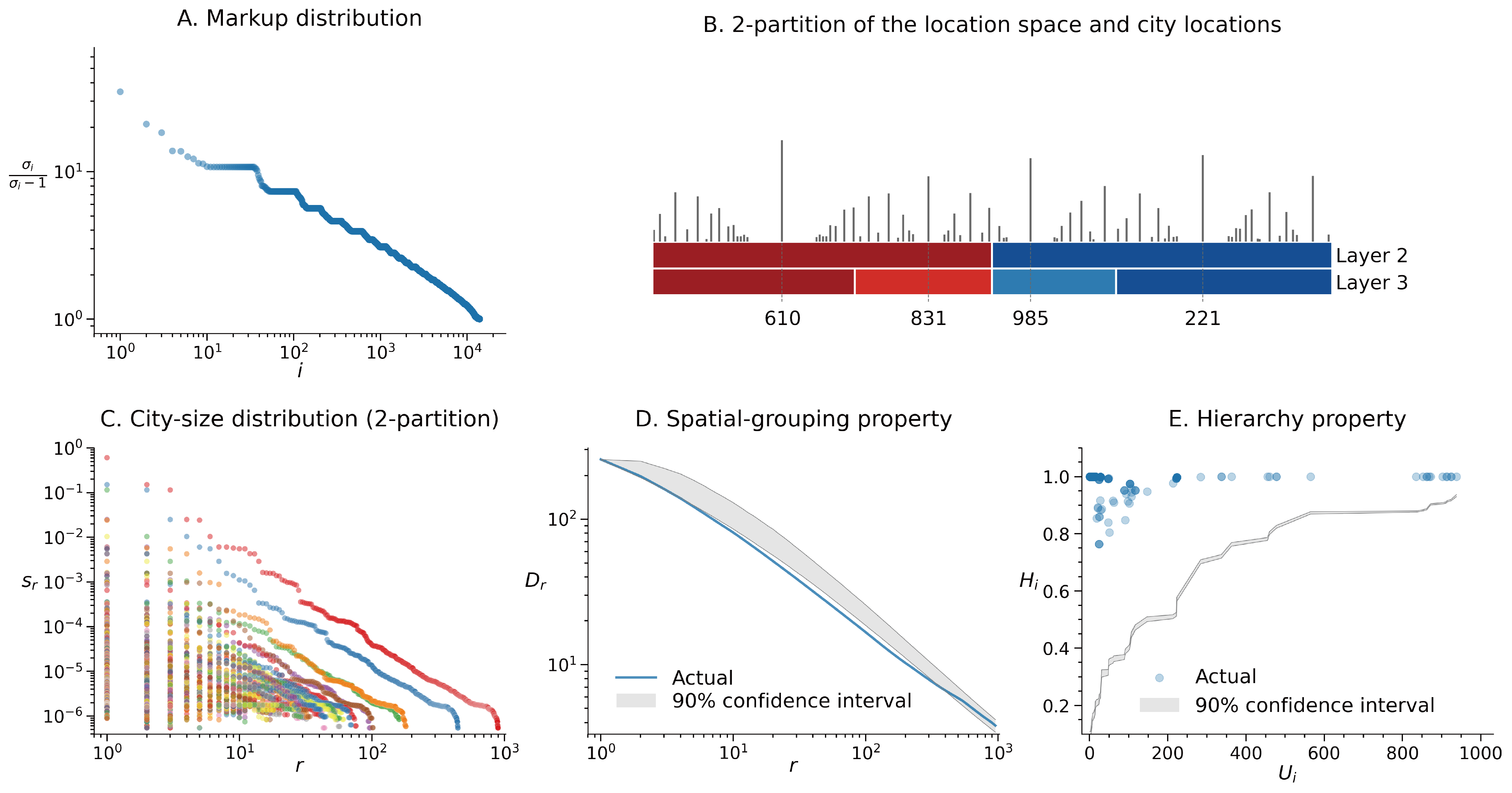}
    \caption{(A) The blue plot shows the distribution of implied price-markup levels (price-marginal cost ratios), $\frac{\sigma}{\sigma_i-1}$, of the 13,930 imported products of the US during 1990--2001 estimated by \protect\cite{Broda-Weinstein-2006}. The products are ordered by the markup level in descending order along the horizontal axis.
   Each orange scatter plot indicates the number of choice cities for the corresponding industry realized in one of the 1,000 equilibrium samples under $(R,I)=(1\textrm{,}024,256)$.
   Panels (B--E) are concerning the first equilibrium sample under $(R,I)=(1\textrm{,}024,256)$. (B) Spatial population distribution with the bar height being the log of city sizes; shown below are the second and third layers of the 2-partition with indication of the locations of four central places. (C) City-size distributions in the 2-partition cells.
    (D,\,E) Results of the SGP and HP tests for the equilibrium sample, corresponding to Fig.~\ref{fig:us}E and F, respectively.}
    \label{fig:sm}
\end{figure*}

\begin{figure*}[h!]
    \centering
    \includegraphics[width=11.4cm]{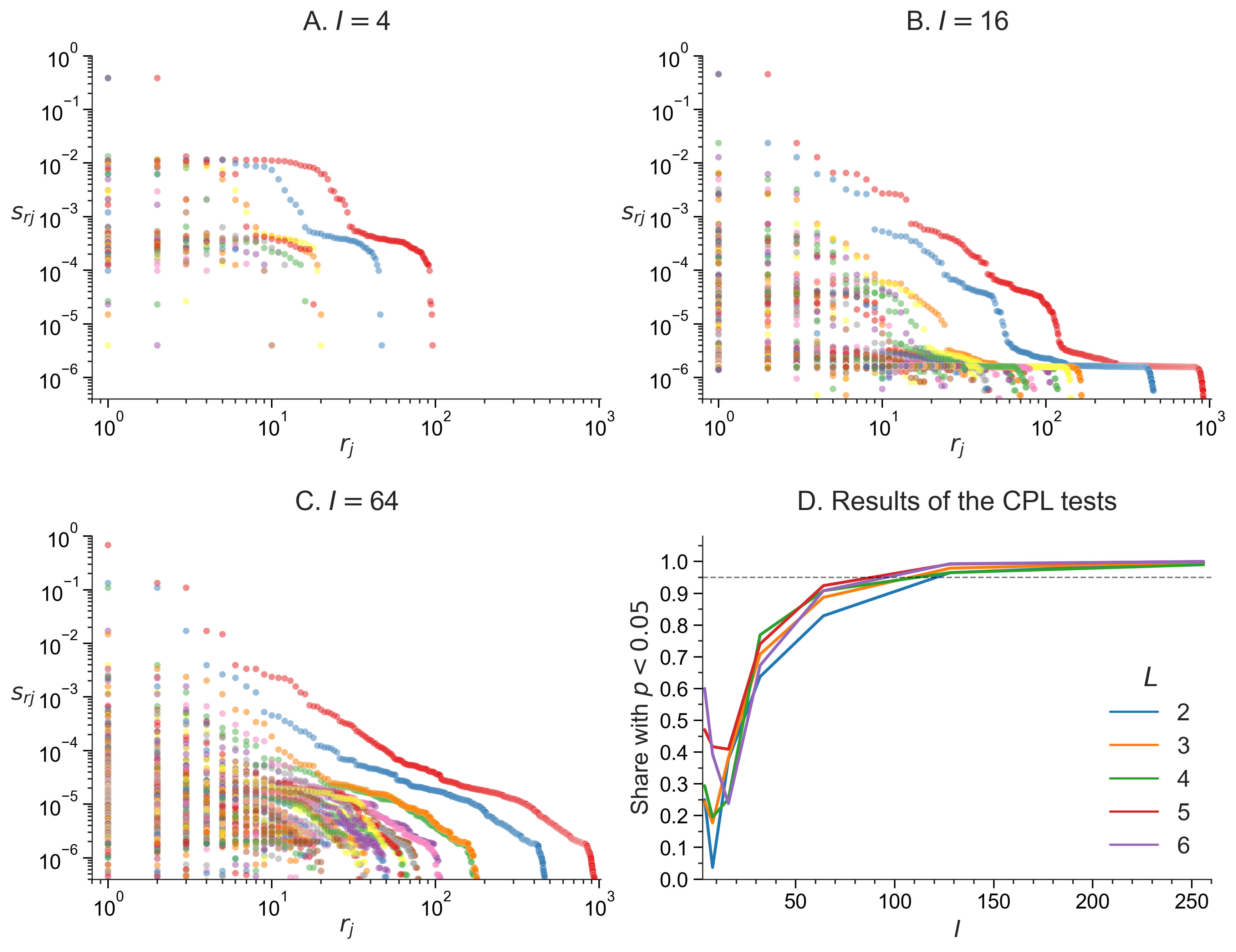}
    \caption{(A--C) City-size distributions in the 2-partition cells of the first equilibrium sample under $I=4, 16, 64$, respectively, and $R=1\rm{,}024$. 
    (D) Share of equilibrium samples under $I=4,8,16,32,64,128,256$ and $R=1\rm{,}024$ for which $p$-values under the null hypothesis of the CPL test are smaller than 0.05.
    ($L=2,\ldots,6$).} 
    \label{fig:cplsm}
\end{figure*}

Second, our simulation results suggest that the regularities emerge in equilibrium under weaker conditions than those derived by \cite{Hsu-2012}, 
which may explain the ubiquity of these regularities in the actual city systems. 
The critical assumption in \cite{Hsu-2012} is that the distribution of the scale economy parameter is scale-free toward the tail.
However, the empirical distribution of scale economies we adopt (Fig.\,\ref{fig:markups}B) does not exhibit scale-freeness toward the tail. Yet, the apparent CPL realizes in most equilibria in our model (Fig.\,\ref{fig:sm}C, Fig.\,\ref{fig:cplsm}D) under a large number of industries.

The regularities identified in this paper impose constraints on feasible urban planning at each regional scale.
The success or failure of place-based policies designed to take advantage of individual cities' characteristics \cite{Neumark-HB2015} depends on their spatial relationships with other cities, which are subject to the self-organized nationwide spatial fractal structure. 
Specifically, the city-size power law constrains the number of cities of a given size or larger. At the same time, the CPH restricts the spatial frequency of these cities and that of industries characteristic to them.
That said, understanding the economic forces behind the formation of the CPH is crucial for successful policymaking.
In particular, in our framework, the origins of the considerable diversity in scale economy parameters are yet to be understood. 

Finally, since our theory considers static aspects of city systems, it is complementary to approaches of dynamic city growth as represented by random growth models \cite{Gabaix-1999,Verbavatz-Berthelemy-Nature2020}. 
Considering extensions of random growth models that add spatial relations among cities \cite{Rybski-etal-PysRevE2013} may be a good starting point for unifying central place theory and dynamic theories. 

%% file: 0-ack.tex
We thank Hiroshi Nakashima for his support in developing the Fortran program for parallel computation.
We benefited from Diego Puga in constructing the central-place property test.
This work was part of the ``Development of Quantitative Framework for Regional Economy based on the Theory of Economic Agglomeration'' project at the Research Institute of Economy, Trade, and Industry. 
T.\,Akamatsu, T.\,Mori, and Y.\,Takayama acknowledge financial support from JSPS Grants-in-Aid for Scientific Research (Grant 17H00987).
T.\,Mori acknowledges financial support from the Kajima Foundations and the Murata Science Foundation. 
This work was supported by the JST FOREST Program (Grant Number JPMJFR215M). 
The numerical simulation analysis in this paper was supported by the ``Joint Usage/Research Center for Interdisciplinary Large-scale Information Infrastructures'' and ``High-Performance Computing Infrastructure'' in Japan (Project ID JH160015-NAH).

%% file: 0-text-si.tex
\setcounter{page}{1}
\setcounter{figure}{0}
\setcounter{equation}{0}

\renewcommand{\thesection}{S\arabic{section}}
\renewcommand{\thepage}{S\arabic{page}}
\renewcommand{\thefigure}{S\arabic{figure}}
\renewcommand{\thetable}{S\arabic{table}}
\renewcommand{\theequation}{S\arabic{equation}}

\section{Materials and Methods\label{sec:si-method}}
\subsection{Data}
\subsubsection*{Population count data} 
Population count data of the United States (US) in years 2000, 2005, 2010, 2015, and 2020, for each 30’’-by-30’’ grid is obtained from the High Resolution Global Population Data Set by the Oak Ridge National Laboratory \cite{LandScan2000,LandScan2005,LandScan2010,LandScan2015,LandScan2020}.
The numbers of cities identified are, 1,224, 1,491, 1,429, 1,439, and 1,344, respectively.

For Japan, population-count data in 30’’-by-45’’ grids are from the Grid Square Statistics of the Census of Japan in 1970, 1975, 1980, 1985, 1990, 1995, 2000, 2005, 2010, 2015, and 2020. The numbers of cities identified are, 504, 511, 496, 483, 484, 477, 472, 471, 448, 442, and 431, respectively (Fig.\,\ref{fig:jp}A).

\subsubsection*{Bilateral road distances}
The road distance between each pair of cities is computed as the shortest-path road distance between the two cities, and the most densely populated grids are chosen to represent the locations of cities. 
The road network data are from OpenStreetMap (\url{http://download.geofabrik.de/}). 
To compute road distances, we follow the instructions given in the Github webpages (\url{https://github.com/Project-OSRM/osrm-backend/wiki/Building-OSRM} and \url{https://github.com/Project-OSRM/osrm-backend/wiki/Building-OSRM}).

\subsubsection*{Industrial locations}

For the case of the US, industrial locations are obtained from the ``Complete ZIP Code Industry Detail File'' from County Business Patterns in 2000 and 2020 of the US Census Bureau (\url{https://www.census.gov/data/datasets/2000/econ/cbp/2000-cbp.html}, \url{https://www.census.gov/data/datasets/2020/econ/cbp/2020-cbp.html}) in which the zip codes of all establishments in up to six-digit NAICS industries are available.
We utilize three-, four-, five- and six-digit levels of industry aggregation for manufacturing, service, wholesale, and retail by the NAICS classification, including 72, 259, 597, and 999 industries, respectively, in 2000, and 75, 260, 561, and 786 industries, respectively, in 2020.
For a given industry aggregation, the set of all the relevant industries is denoted by $\bm I$.
We associate establishments with cities by matching each zip-code polygon to the city polygon, which accounts for the largest areal share in the zip-code polygon.

For the case of Japan, we focus on the same categories of manufacturing, service, wholesale, and retail but using the three-digit Japanese Standard Industrial Classification (JSIC) obtained from the Establishment and Enterprise Census in 2001 and the Economic Census for Business Frame in 2014.
They are matched to cities in 2000 and 2015, respectively. 
Coordinates of establishment locations are matched to the city polygon.

\subsubsection*{Substitution elasticities in the simulation}
The substitution elasticities of products used in the simulations are drawn from those of the 13,930 imported products of the US in the 1990--2001 period estimated by \cite{Broda-Weinstein-2006}.
The products are classified according to the 10-digit Harmonized Tariff System (HTS).
The source data are available from
\url{http://www.columbia.edu/~dew35/TradeElasticities/TradeElasticities.html}.
\bigskip

\noindent\dataset{Dataset.pdf, Simulation.7z, Results.7z.}\label{dataset}~Fortran programs for simulations, associated input data, equilibrium samples under $R=1024$ and $I=2, 4,8, 16,32, 64, 128$ and $256$, as well as Python programs and the processed data underlying the numbers and figures are available at  \url{https://www.dropbox.com/sh/a1riz3rz17urnht/AACpTZt5AaLpBN60JVwgXFgZa?dl=0}.

\subsection{Hypothesis tests}

For a given test statistic $x$, let $x_0$ and $\tilde{x}$ be the values implied by the actual and counterfactual data, respectively. 
Our null hypothesis is then given by 
\begin{quote}
\emph{$H_0$: $x_0$ and $\tilde{x}$ belong to the same statistical population.}    
\end{quote}

We generate $M$ random counterfactual values of the test statistic, $\tilde{x}_m$  for $m=1,\ldots,M$.
Assuming the observed value $x_0$ is also from the statistical distribution under $H_0$, the effective sample size under $H_0$ is $M+1$. If $M_0$ denotes the number of instances $x_m$ which are at least as large (or small) as the observed $x_0$, then the $p$-value, $p_0$, for a one-sided test of $H_0$ is given by 
\begin{equation}
    p_0 = \frac{M_0}{M+1},\label{eq:p-value}
\end{equation}
where $M=1\rm{,}000$ in all our tests. 
For example, if from among these samples say $30\left(=M_{0}\right)$ are as large (or small) as the observed value, then under $H_{0}$ the chance of observing a value this large (small) is $p_{0}=30/1001\approx 0.03$, which provides substantial evidence against $H_{0}$.

In the cases of the CPL, SGP and HP tests, the test statistics are given by MSE, $D_r$ (Eq.\,\ref{eq:SGP}) and $H_i$ (Eq.\,\ref{eq:h}), respectively.
The relevant tests are one-sided for all the tests.
The CPL implies $\textrm{MSE} < \widetilde{\text{MSE}}$, the SGP implies $D_r < \tilde{D}_r$, and the HP implies $H_i > \tilde{H}_i$. %

\subsection{Theoretical model}
We consider an economy with a circular geography having equi-spaced discrete locations, $\bm R = \{1,2,\ldots,R\}$, along the circumference of a circle of a unit radius. 
There is a continuum of mobile and immobile workers of given sizes $H$ and $L$, respectively, where each worker is endowed with a unit of the corresponding labor.
Mobile workers freely choose the location where they reside and work, while immobile ones are uniformly distributed in $\bm R$ so that $l = L/R$ units of immobile workers are attached to each location.

There are two sectors, rural and urban, in this economy.
Rural sector is perfectly competitive, and produces a homogeneous
good subject to the constant-returns technology.
Urban sector consists of $I$ monopolistically competitive industries,
each producing a variety of differentiated goods subject to the firm-level increasing-returns technology.

For simplicity, we assume that only urban products are subject to transportation costs. 
Specifically, in order for a unit of good to be supplied from location $r$ to $s\neq r$, $\tau_{ij}>1$ units must be shipped from $r$ so that $\tau_{ij}-1$ units melt down en route. Shipments within a location are assumed to be costless so that $\tau_{ii}\equiv 1$.

\subsubsection*{Consumer behavior}
In this economy, consumers are workers. Their preference is given by the quasi-linear utility function,
\begin{equation}
    W (\{C_i\}_{i\in\bm I},A) = \sum_{i\in\bm I} \ln C_i + A,
\end{equation} 
where $A$ represents the consumption of rural products,
\begin{align}
C_i\equiv\left( \int_0^{n_i} q_i(\omega)^\frac{\sigma_i - 1}{\sigma_i} d\omega\right)^\frac{\sigma_i}{\sigma_i-1}
\end{align} 
is a composite of $n_i$ varieties of industry $i$'s product with $q_i(\omega)$ being the consumption density of variety $\omega\in [0,n_i]$. 
Substitution elasticity between each pair of product-$i$ varieties is given by a constant, $\sigma_i>1$, which means that each variety is not necessary, and the mass of varieties $n_i$ is endogenous.

Assuming symmetry among the product varieties of the same industry supplied from the same location, the industry-$i$ composite consumed in location $r$ can be written as
\begin{equation}
    C_{ir} = \left( \sum_{s\in\bm R} n_{i,s} q_{i,sr}^\frac{\sigma_i - 1}{\sigma_i}\right)^\frac{\sigma_i}{\sigma_i-1},
\end{equation}
where $n_{i,s}$ is the mass of industry $i$'s varieties produced in location $s$, and $q_{i,sr}$ is the consumption of each of these varieties supplied from $s$.

Let the rural product be num\'{e}raire, consumers' utility maximization under given nominal income, $Y$, implies the optimal demands in each location $r\in \bm R$,
\begin{align}
    A &= Y - I,\\
    C_{ir} &= 1/P_{ir},
\end{align}
where $P_{ir}$ represents the price index of industry $i$'s product-variety composite in location $r$, which by denoting the delivered price of $i$'s variety produced in location $s$ and consumed in location $r$ by $p_{i,sr}\equiv p_{is}\tau_{rs}$, is expressed as
\begin{equation}
    P_{ir} = \left( \sum_{s\in\bm R} n_{ir} p_{i,sr}^{1-\sigma_i}\right)^\frac{1}{1-\sigma_i}.\label{eq:price-index}
\end{equation}
The individual demand for industry $i$'s variety produced in location $s$ and consumed in $r$ is given by
\begin{equation}
    q_{i,sr} = p_{i,sr}^{-\sigma_i}P_{ir}^{\sigma_i-1}.\label{eq:demand}
\end{equation}

\subsubsection*{Producer behavior}
Given a potentially infinite mass of varieties assumed to exist for each product, each firm produces a unique variety, that is, $n_{ir} = h_{ir}\,\forall (i,r)\in \bm I\times\bm R$, since the production is subject to the firm-level scale economies.
In monopolistically competitive markets, firms set monopolistic prices for their output, while their profit is driven down to zero in equilibrium under free entry and exit.
The measure $n_i$ of product variety is thus determined endogenously.

The production of industry $i$'s variety requires a unit of mobile labor as fixed input and $\beta$ units of intermediate good for every unit of output. 
Each unit of intermediate and rural  goods is produced using one unit of immobile labor; hence, the wage rate of immobile labor is 1.
The cost for producing amount $x$ of the output is given by
\begin{equation}
    c(x,w) = w + \beta x,
\end{equation}
where $w$ is the wage rate of mobile labor.
The monopolistic prices of industry $i$'s variety is given by
\begin{equation}
    p_{i} = \frac{\sigma_i}{\sigma_i - 1} \beta \label{eq:price}
\end{equation}
irrespective of their location.
Notice that the price markup, $\frac{1}{\sigma_i-1}$, over the marginal cost, $\beta$, is larger for more differentiated products (with smaller $\sigma_i$). 

\subsubsection*{Short-run equilibrium}
We consider short- and long-run adjustments out of equilibria.
In the short run, the distribution of mobile workers across locations and industries is given. Consumers maximize their utility, firms earn zero profit  at optimum due to free entry and exit, and all markets clear in equilibrium.

Let $\bm h = \{h_{ir}\}_{r\in\bm R, i\in\bm I}$ such that $\sum_{i\in\bm I, r\in\bm R} h_{ir} = H$ is a distribution of mobile workers in a short-run, where $h_{ir}$ is the population size of mobile workers employed in industry $i\in \bm I$ in location $r\in\bm R$.

The total demand in location $r$ for any product-$i$ variety produced in location $s$ is given by 
\begin{equation}
    Q_{i,sr} = q_{i,sr} (h_r + l),\label{eq:total-demand}
\end{equation}
where $h_r = \sum_{i\in\bm I} h_{ir}$ is the population size of mobile workers in location $r$.
In short-run equilibrium, the market for each product variety clear, so that 
\begin{equation}
    x_{ir} = \sum_{s\in\bm R} \tau_{rs} Q_{i,sr}.\label{eq:market}
\end{equation}

Let the \emph{freeness-of-trade measure} defined by 
\begin{equation}
   \phi_{i,sr} \equiv \tau_{sr}^{1-\sigma_i}.
\end{equation}
Then, the \emph{market acessibility} to product-$i$ varieties from location $r$ can be expressed by
\begin{equation}
    \Delta_{ir} (\bm h_i) \equiv \sum_{s\in\bm R} \phi_{i,sr} h_{is},\label{eq:access}
\end{equation}
where $\bm h_i \equiv (h_{i1},\ldots,h_{iR})^\top$ is a vector of industry-$i$ employment across locations.
Since the production of each variety requires a unit of mobile labor, we have $h_{ir} = n_{ir}$.
Then, by substituting Eq.\,\ref{eq:price} into Eq.\,\ref{eq:price-index}, and using Eq.\,\ref{eq:access}, the price index of product-$i$ varieties can be rewritten as
\begin{equation}
    P_{ir} (\bm h_i) = \frac{\sigma_i}{\sigma_i - 1}\beta \Delta_{ir} (\bm h_i)^\frac{1}{1-\sigma_i}.\label{eq:price-index-2}
\end{equation}

Due to the zero profit under free entry and exit, the wage rate of mobile labor for industry $i$ in location $r$ is obtained as
\begin{equation}
    w_{ir} = (p_{ir} - \beta) x_{ir}.\label{eq:zero-profit}
\end{equation}
By substituting Eqs.\,\ref{eq:market} and \ref{eq:price} into Eq.\,\ref{eq:zero-profit}, and using Eqs.\,\ref{eq:price-index-2}, \ref{eq:demand} and \ref{eq:total-demand}, the wage rate for mobile workers in location $r$ is obtained as
\begin{equation}
    w_{ir} (\bm h_{i}) = \frac{1}{\sigma_i} \sum_{s\in\bm R} \frac{\phi_{sr}}{\Delta_{is} (\bm h_i)} (h_s + l).
\end{equation}
The total income of mobile workers in location $r$ can be expressed by
\begin{equation}
    w_{ir} h_{ir} = \frac{1}{\sigma_i} \sum_{s\in\bm R} A_{i,sr} (h_s + l),
\end{equation}
where
\begin{equation*}
    A_{i,sr} \equiv \frac{\phi_{i,sr} h_{ir}}{\Delta_{ir} (\bm h_i)}
\end{equation*}
is the share of location $r$ in the product-$i$  market in location $s$, where share $1/\sigma_i$ of the total sales is paid to mobile workers, while $1-1/\sigma_i$ is payed to immobile ones.

\subsubsection*{Long-run equilibrium and its stability}
In the long-run, mobile workers migrate across locations and change industries to be employed, seeking higher utility levels. 
A long-run equilibrium is a short-run equilibrium in which mobile workers (and firms) have no incentive to relocate.

Define the $R\times R$ \emph{spatial discounting matrix}, $\mathbf{D}_i$,
for industry $i\in\bm I$ whose element $(r,s)$ is given by $\phi_{i,rs}$,
then the vector of (short-run) market accessibilities given $\bm h_i$ in locations defined by Eq.\,\ref{eq:access} can be written as
\begin{gather}
\mathbf{\Delta}_i\equiv\left(\Delta_{i1}(\bm h_i),\ldots,\Delta_{iR}(\bm h_i)\right)=\mathbf{D}_i^\top\bm h_i\:.\label{eq:market-access-matrix}
\end{gather}
Using this, the short-run equilibrium utility levels, $\bm v_i (\bm h) \equiv [v_{i1} (\bm h), \ldots, v_{iR} (\bm h)]^\top$, for $(i,r)\in \bm I\times \bm R$ are solved in a closed form given the distribution of mobile workers, $\bm h$, as
\begin{equation}
\bm v_i(\bm h)=\sum_{j\in\bm I}\bm S_j(\bm h_j)+\bm w_i (\bm h)-I\mathbf{1}\label{eq:indirect-utility}
\end{equation}
where $\bm S_i (\bm h_i) \equiv [S_{i1}(\bm h_i),\ldots,S_{iR} (\bm h_i)]^\top$ is the vector of industry-$i$ market accessibility from each location of consumers, and $\bm w_i (\bm h) \equiv [w_{i1} (\bm h),\ldots,w_{iR} (\bm h)]^\top$ is the vector of wage rates for mobile workers employed in industry $i$ in each location, given respectively by
\begin{gather}
\bm S_i(\bm h_i)\equiv(\sigma_i-1)^{-1}\ln[\mathbf{D}_i\bm h_i],\label{eq:good-accessibility}\\
\bm w_i(\bm h)\equiv\mathbf{D}_i\left(\text{diag}\mathbf{\Delta}_i\right)^{-1}(\bm h +l\mathbf{1}),\label{eq:wage-matrix}
\end{gather}
Here, ${\bm h}\equiv[h_{1},\ldots,h_{R}]^{\top}$, and $\mathbf{1}$
is a $R\times1$ vector of which all elements are one; $\ln[\mathbf{a}]\equiv[\ln a_{1},\ln a_{2},\ldots]^\top$. 

The conditions for the long-run equilibrium are given by
\begin{subequations}
\begin{gather}
\begin{cases}
v^{*}-v_{ir}(\bm h)=0\;, & \text{if}\;h_{ir}\geq 0\\
v^{*}-v_{ir}(\bm h)>0\;, & \text{if}\;h_{ir}=0
\end{cases}\quad\forall (i,r)\in\bm I\times \bm R\,,\label{eq:long-run-a}\\
\sum_{i\in\bm I, r\in\bm R} h_{ir}=H\,,\label{eq:long-run-b}
\end{gather}
\label{eq:long-run}\end{subequations}where $v^{*}$is the long-run equilibrium utility level of mobile workers. 

It is well known that the solution to the non-linear complementarity problem, such as that in Eq.\,\ref{eq:long-run-a} can be rewritten
as the fixed point of the projection dynamics,
\begin{gather}
\dot{\bm h}(t)=\bm F (\bm h(t))-\bm h(t)\;,\label{eq:adjustment}
\end{gather}
where $t$ represents the fictitious time along which the migration
of mobile workers take place, and $\bm F (\bm h)$ is the
projection operator, $\bm F(\bm h) = \text{Proj}_{\Omega}\left(\bm h+\bm v(\bm h)\right)$, defined by
\begin{equation}
\text{Proj}_{\Omega}[\bm x]\equiv\arg\min_{\bm z}\left\{ (\bm z-\bm x)\cdot(\bm z-\bm x)\;\text{s.t.}\;\bm z\in\Omega\right\} \;,\label{eq:projection}
\end{equation}
where $\Omega$ is the $(I\times R)$-dimensional simplex defining feasible distributions of mobile workers across locations and industries given by 
\begin{equation}
\Omega\equiv\Bigg\{ \bm h\,\Bigg|\sum_{(i,r)\in\bm I\times\bm R} h_{ir}=H,\;h_{ir}\geq 0\Bigg\} \;.\label{eq:feasible-worker-distribution}
\end{equation}

\subsection{Simulation procedure}
To solve our high-dimensional and highly non-linear problem, we adopt the \emph{merit function approach} \cite{Fukushima-MP1992}. 
This approach utilizes the equivalence between the non-linear complementarity problem
given by Eq.\,\ref{eq:long-run} and the variational inequality problem given by
\begin{equation}
\text{Find}\;\bm h^* \in\Omega\;\;\text{s.t.}\;-\bm v (\bm h^*)\cdot(\bm h-\bm h^*)\geq0\quad\forall\,\bm h\in\Omega\;,\label{eq:variational-inequality-prob}
\end{equation}
where ``$\cdot$'' here means an inner product.

Fukushima \cite{Fukushima-MP1992} proposes to solve (\ref{eq:variational-inequality-prob})
by means of the following minimization problem,
\begin{equation}
\min_{\bm h\in\Omega}G(\bm h)=-\bm v(\bm h)\cdot\left(\bm h-\bm F(\bm h)\right)-\frac{1}{2}(\bm F(\bm h)-\bm h)\cdot (\bm F (\bm h)-\bm h)\:.\label{eq:merit-function}
\end{equation}
Here, $G(\bm h)$ is called the merit function, and it has been shown by \cite{Fukushima-MP1992} that if $\bm h$ is a solution to the problem (\ref{eq:variational-inequality-prob}), it must hold that $G(\bm h)=0$;
otherwise, $G$ is a strictly positive function.

In the simulations, we follow the steps below to identify the long-run equilibrium starting
from a given initial distribution of mobile workers. 
\medskip

\begin{minipage}[t]{0.9\hsize}%
\begin{enumerate}
\item[Step 1.] Set the initial distribution of skilled workers, $\bm{h}^{(1)}\in\Omega$.
\item[Step 2.] Set the direction of adjustment:
\[
\bm{d}^{(n)}\coloneqq\bm{F}(\bm{h}^{(n)})\;.
\]

\item[Step 3.] Determine the step size based on \emph{Armijo} \emph{rule}: Let $\delta$
and $\gamma$ be given constants such that $\delta>0$ and $0<\gamma<1$. 
For the current iterate $\bm{h}^{(n)}$ and the search direction $\bm{d}^{(n)}$, determine the step length $\gamma^{k(n)}$ where
$k(n)$ is the smallest non-negative integer $k$ such that
\[
G(\bm{h}^{(n)}+\gamma^{k}\bm{d}^{(n)})- G(\bm{h}^{(n)}) \leq -\delta\gamma^{k(n)}\left\Vert \bm{d}^{(n)}\right\Vert ^{2},
\]
where $\left\Vert \bm x\right\Vert $ represents the Euclid norm of a
vector $\bm x$ (see Theorem 4.2 of \cite{Fukushima-MP1992} for
the detail).

\quad If 
$\gamma^{k(n)}>\gamma_{\min}$, update the solution as $\mathbf{h}^{(n+1)}\coloneqq\mathbf{h}^{(n)}+\gamma^{k(n)}\mathbf{d}^{(n)}$.
Otherwise, update the solution as
$\mathbf{h}^{(n+1)}\coloneqq\mathbf{h}^{(n)}+\bar{\gamma} \mathbf{d}^{(n)}$, where $\bar{\gamma}>0$ is a given constant.
\item[Step 4.] If %
$G(\bm{h}^{(n+1)})<10^{-8}$, stop. 
Otherwise, let $n\coloneqq n+1$, and move to Step 1.\end{enumerate}
\end{minipage}

\bigskip{}
\bigskip{}

If the objective function $G(\cdot)$ is strictly convex and has a unique solution, then the solution update given in Steps 3 and 4 will eventually reach the solution. 
However, generally $G$ is not convex and has local minima such that $G(\bm{h})>0$. 
The perturbation of $\bm{h}$ toward the direction of the projection dynamics (\ref{eq:adjustment})
enables us to escape from these local minima.

In the simulation, we set 
\begin{align*}
&\delta=10^{-5}, \gamma=0.8, \gamma_{\min}= 5.0\times  10^{-3-\min\{5, 10^{-6}n\}}, \\
&\bar{\gamma}= 
\begin{cases}
    5.0\times 10^{-3} &\text{if}\quad G(\mathbf{h}^{(n)}>10^{-6},
    \\
    2.0\times 10^{-2} &\text{otherwise}.
\end{cases} 
\end{align*}

\subsection{Diversity in scale economies}\label{app:markup}
\begin{figure*}[h!]
    \centering
    \includegraphics[width=0.8\hsize]{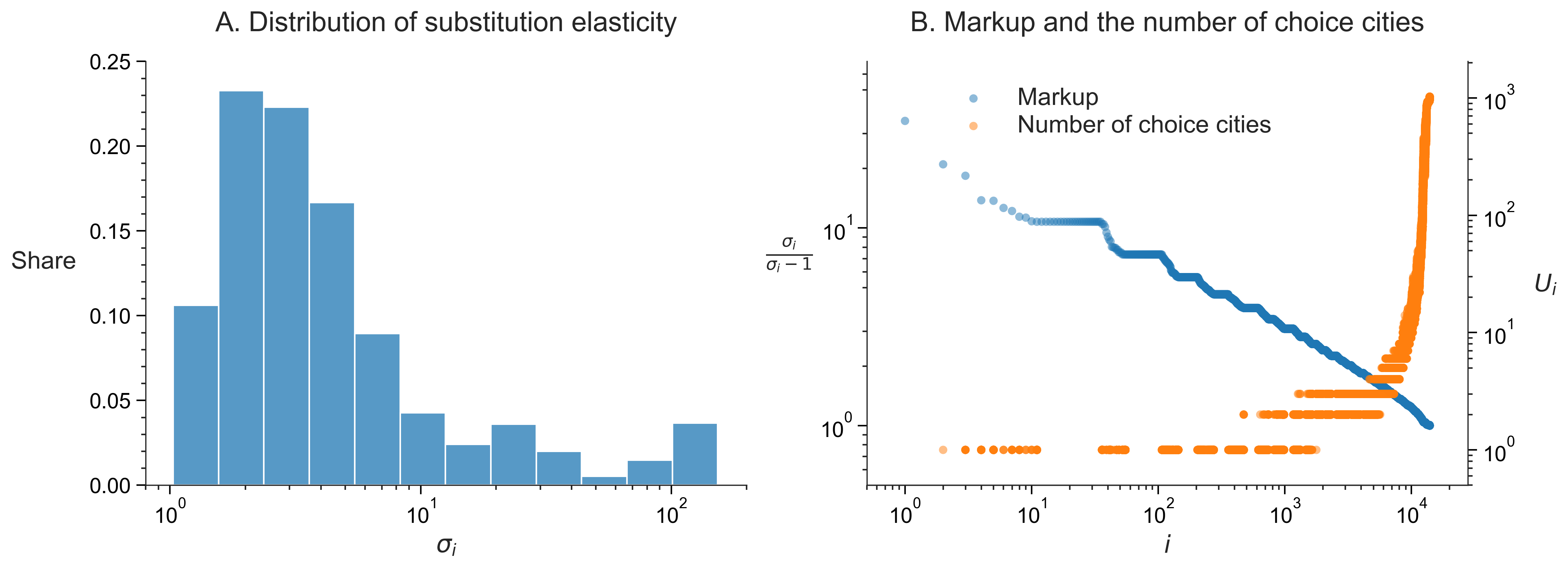}
   \caption{(A) Frequency distribution of substitution elasticities of the 13,930 imported products (according to the 10-digit HTS classification) in the US during 1990--2001 estimated by \protect\cite{Broda-Weinstein-2006}. (B) The blue plot shows the distribution of implied price-markup levels under these substitution elasticities. The products $i$ are ordered by the size of markup as the ratio of f.o.b. price over marginal cost, $\frac{\sigma}{\sigma_i-1}$, in descending order.
   Each of the orange scatter plots indicates the number of choice cities for the corresponding industry in an equilibrium in one of the 1,000 equilibrium samples under $(R,I)=(1024,256)$.} 
    \label{fig:markups}
\end{figure*}

Substitution elasticities, $\sigma_i$, for a given industry in the model are drawn from set $\boldsymbol\Sigma$ consisting of 13,930 values of the estimated substitution elasticities of the imported products of the US in the 1990--2001 period \cite{Broda-Weinstein-2006}. 
The products are classified according to the 10-digit HTS.
Fig.\,\ref{fig:markups}A shows the empirical frequency distribution of $\sigma$ values. 
The blue plots in Fig.~\ref{fig:markups}B show the values of price-markup $p_i/\beta = \frac{\sigma_i}{\sigma_i - 1}$ implied by our model (Eq.\ref{eq:price}) against their ranking $i$ in descending order.  
The markup values are diverse, ranging from 1.0002 ($\sigma_i = 4\rm{,}303$) to 34.33 ($\sigma_i=1.03$), with a median of 1.467 ($\sigma_i = 3.14$).

Notice not only their wide range but also the dense intermediate values.
The industry-specific price-markup level reflects the degree of scale economies of the industry, and thus, reflects their tendency of spatial concentration.
For illustration, each of the orange scatter plots in Fig.\ref{fig:markups}B indicates the equilibrium number of choice cities for an industry in the HTS system associated with the corresponding $\sigma_i$-value in one of the 1,000 simulated equilibrium samples under $(R,I)=(1024,256)$. 
The number of choice cities for an industry, $U_i$, is largely dependent of the markup level $\frac{\sigma_i}{\sigma_i - 1}$. 
Let $m_U$ be the threshold value of the markup 
where all industries with $\frac{\sigma_i}{\sigma_i - 1} \ge m_U$ never locate in more than $U$ cities for any of the $1,000$ computed samples. 
The threshold values are obtained as 
$m_2 = 2.9$, 
$m_4 = 1.57$, 
$m_8 = 1.34$, 
$m_{16} = 1.25$, 
$m_{32} = 1.17$, 
$m_{64} = 1.10$, 
$m_{128} = 1.07$,
$m_{256} = 1.05$, and 
$m_{512} = 1.04$, 
highlighting a positive association between the markup level and localization of an industry.

The properties of the distribution of scale economies is known to be crucial for the emergence of the city-size power law in central place theory-inspired models. 
For a stylized central place theory model, Hsu \cite{Hsu-2012} obtained a theoretical characterization, albeit model-specific, for the scale economy distribution under which the city-size distribution satisfies the power law. 
Although Hsu's theory is not directly applicable to our model, the essential idea is that the magnitudes of scale economies (in his framework) are scale-free distributed toward tail, ensuring a large diversity in industrial localization levels. 
From the orange dots in Fig.\ref{fig:markups}B, we observe that the number of industry choice cities in our computed examples is dense in the possible range, $1$ to $1,024$. 
Thus, the emergence of the CPL in our framework may be loosely related to the theoretical finding by Hsu.

\section{Supplementary Text\label{sec:supp-text}}
\subsection{Supplementary results for the United States\label{sec:si-us}}
\subsubsection*{Power laws}
\begin{figure*}[h!]
    \centering
    \includegraphics[width=.8\hsize]{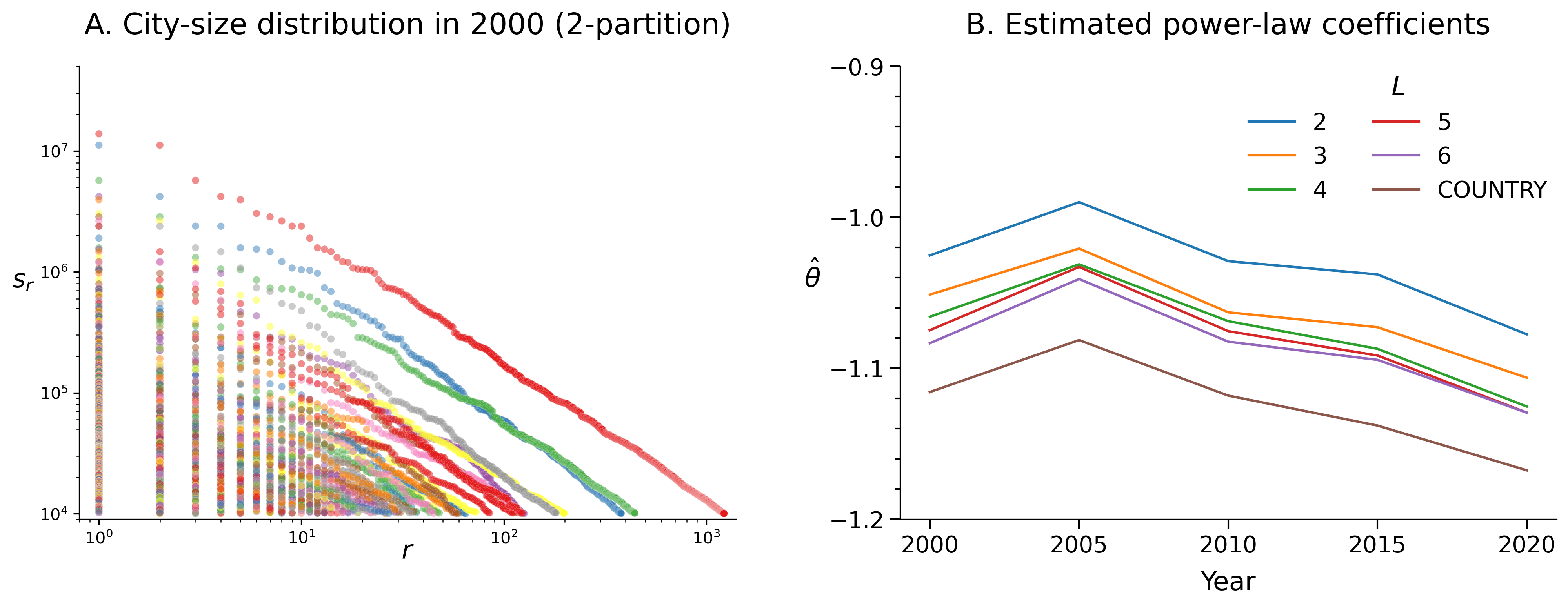}
    \caption{(A) City-size distributions for the 2-partition cells of the (continental) US in 2000.    (B) Estimated power-law coefficients, $\hat{\theta}$ under $L$-partition ($L=2,\ldots,6$) together with that for all cities in the US for years 2000, 2005, 2010, 2015, and 2020.}
     \label{fig:cpl-slopes}
\end{figure*}

City-size distributions of the US cities in 2000, 2005, 2010, 2015, and 2020 are plotted in Fig.\,\ref{fig:cpl-slopes}A.
Despite the volatility in relative city sizes shown in Fig.\,\ref{fig:us}(G,E), the city-size power law at the country level has been stable for the past twenty years.
Fig.~\ref{fig:cpl-slopes}B shows the city-size distributions in the 2-partition cells in 2000.
All the CPL tests under $L=2,\ldots,6$ for years 2000-2020 reject the null hypothesis at the 1\% level.
While the estimated power-law coefficients, $\hat{\theta}$, have become larger in absolute value since 2005 for all $L$, the changes are relatively small (Fig.~\ref{fig:cpl-slopes}).

\subsubsection*{Changes in industrial location between 2000 and 2020}
Table \ref{tb:industry-churning}A (B) lists the ten industries that experienced the largest net decrease (increase) in the number of choice cities between 2000 and 2020.
A large decrease in the number of choice cities in this period reflects the influence of the internet. 
For example, many actual stores of ``Book Stores and News Dealers,'' ``Florists,'' and ``Travel Arrangement and Reservation Services'' are replaced by online services (see \cite{DeAngelis-CNBC2013} for the case of florists).

Most industries exhibiting the largest net increase in the number of choice cities are the ubiquitous industries found in most cities. In many cases, their exits from cities are due to relocations from disappeared cities to newly formed cities.
Ten industries listed in Table \ref{tb:industry-churning}B, for example, are found in more than 95\% of all cities in 2020.

\subsubsection*{Spatial-grouping property}
\begin{figure*}
    \centering
    \includegraphics[width=.4\hsize]{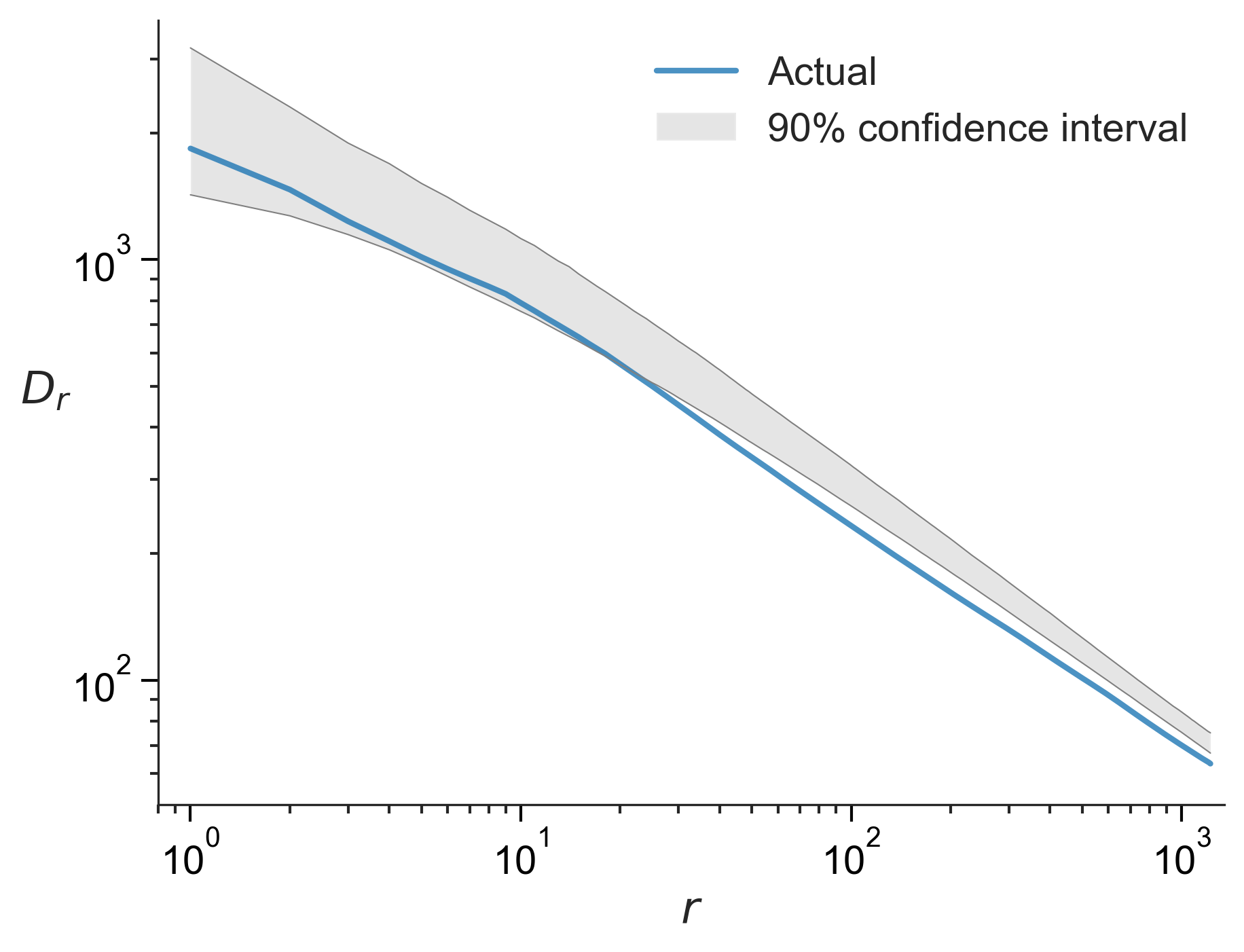}
    \caption{The results of the spatial-grouping property test for the US cities in 2000. The blue curve shows the actual $D_r$ for each $r$, and the shaded area indicates the 90-percent confidence interval of the counterfatual $\tilde{D}_r$ under the null hypothesis.}
     \label{fig:sgp-2000_us}
\end{figure*}

Fig.~\ref{fig:sgp-2000_us} shows the SGP-test result for the US cities in 2000.
The SGP is significant at the 5\% (10\%) level for $r\geq 21, 18, 17$ and $44$ in 2000, 2005, 2010, and 2015, respectively.

\subsubsection*{Central-place property}
Note that $L$-partitions subsume the SGP to hold. 
To evaluate the relevance of the central place theory, we can also test the consistency of an $L$-partition with the SGP.
If the SGP holds for all city sizes, it implies that, given an $L$-partition, each central place in a higher layer must be larger in size than those in a lower layer. 
We call this property the \emph{central-place property} (\emph{CPP}). 

For a given pair of central places $u$ and $u'$, its consistency with the CPP can be expressed by the indicator:
\begin{equation}
    \textit{cp}_{uu'} = 
    \begin{cases}
        1, & \text{if $s_u\geq s_{u'}$ and $\ell_u\leq \ell_{u'},\; u,u'\in \bm U^*_L$}\\
        0, & \text{otherwise,}
    \end{cases}
\end{equation}
where $\bm U^*_L$ is the set of central places in the $L$-partition, and $\ell_u$ is the highest layer in which city $u$ is a central place.

The average consistency of all pairs of larger and smaller central places with the CPP is then given by
\begin{equation}
    \textit{cp}_L = \frac{1}{U_L^*- (L + 1)} \sum_{(u,u')\in \bm U_L^*\backslash \bm U_{L+1}} \textit{cp}_{uu'} \in [0,1].
\end{equation}
The CPP is always satisfied for the set $\bm U_{L+1}$ of the $L+1$ central places in the first and second layers; the two layers are excluded in the computation of $\textit{cp}_L$.

The random counterfactual values, $\widetilde{\textit{cp}}_L$, of $\textit{cp}_L$ are computed for the counterfactual $L$-partition constructed for the CPL test. 
The null hypothesis to test is that $\textit{cp}_L$ and the counterfactual $\widetilde{\textit{cp}}_L$ belong to the same statistical population.

\begin{figure*}
    \centering
    \includegraphics[width=.7\hsize]{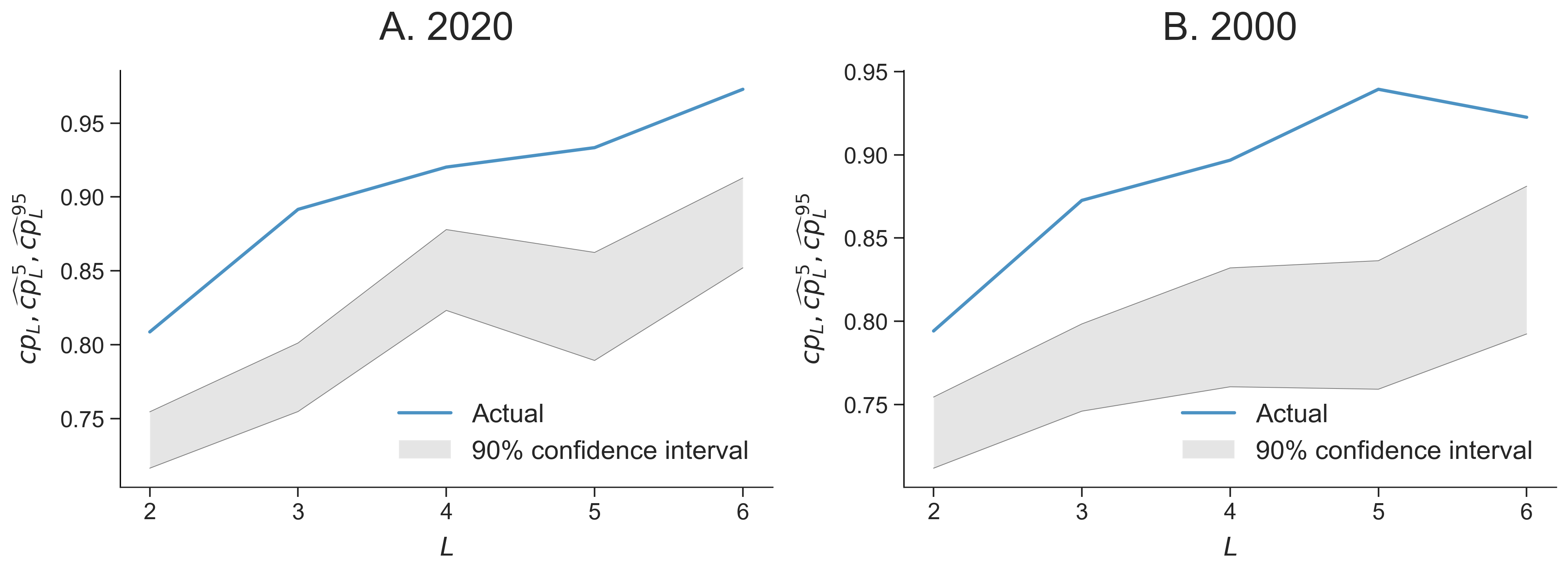}
    \caption{(A,\,B) The results of the central-place property test for the US cities in 2020 and 2000, respectively, showing the actual $\textit{cp}_L$ (blue curves) and the 90-percent confidence intervals (shaded areas) of the counterfactual $\widetilde{\textit{cp}}_L$ under the null hypothesis.}
     \label{fig:cp_us}
\end{figure*}

Fig.\,\ref{fig:cp_us}A and B show $\textit{cp}_L$ values ($L=2,\ldots,6$) in 2020 and 2000, respectively, with 90\% confidence intervals based on $\widetilde{\textit{cp}}_L$ values. 
Each $\textit{cp}_L$ is persistently larger than their counterfactual values, supporting the CPP and hence the SGP. 
In all $L=2,\ldots,6$ in all years, $2000, 2005,\ldots,2020$, the $p$-values under the null hypothesis are zero.

\subsubsection*{Hierarchy property} 
\begin{figure*}
    \centering
    \includegraphics[width=.8\hsize]{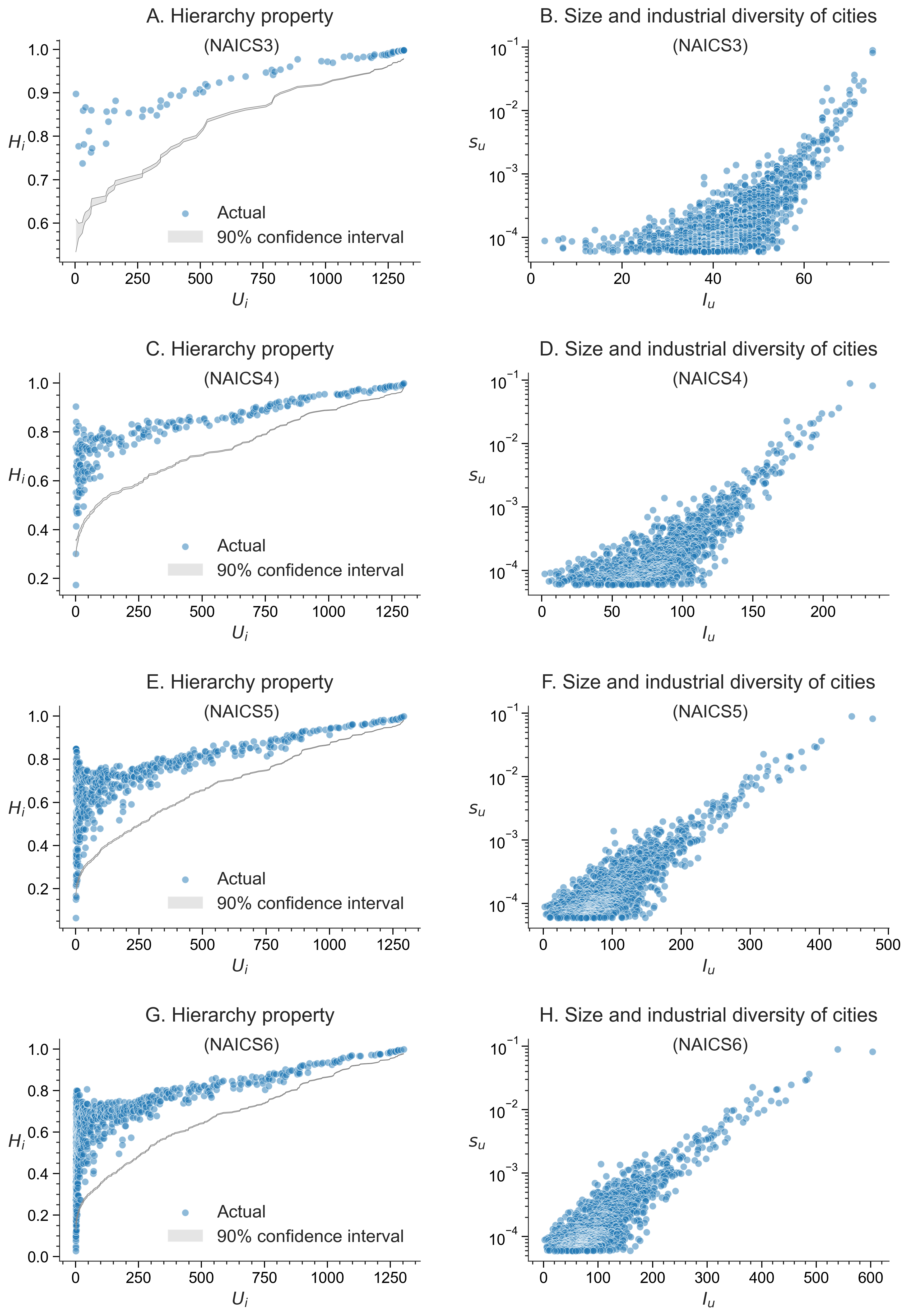}
    \caption{(A,\,C,\,E,\,G) Hierarchy shares $H_i$ plotted against the number $U_i$ of choice cities for the secondary and tertiary industries in the three-, four-, five- and six-digit NAICS, which include 75, 260, 561, and 786 industries, respectively, in 2020.
    The shaded area indicates the 90\% confidence interval of the counterfactual $\tilde{H}_i$ under the null hypothesis. 
    (B,\,D,\,F,\,H) The size $s_u$ and industrial diversity $I_u$ of cities $u$.}
    \label{fig:us_hp_2020}
\end{figure*}

\begin{figure*}
    \centering
    \includegraphics[width=.8\hsize]{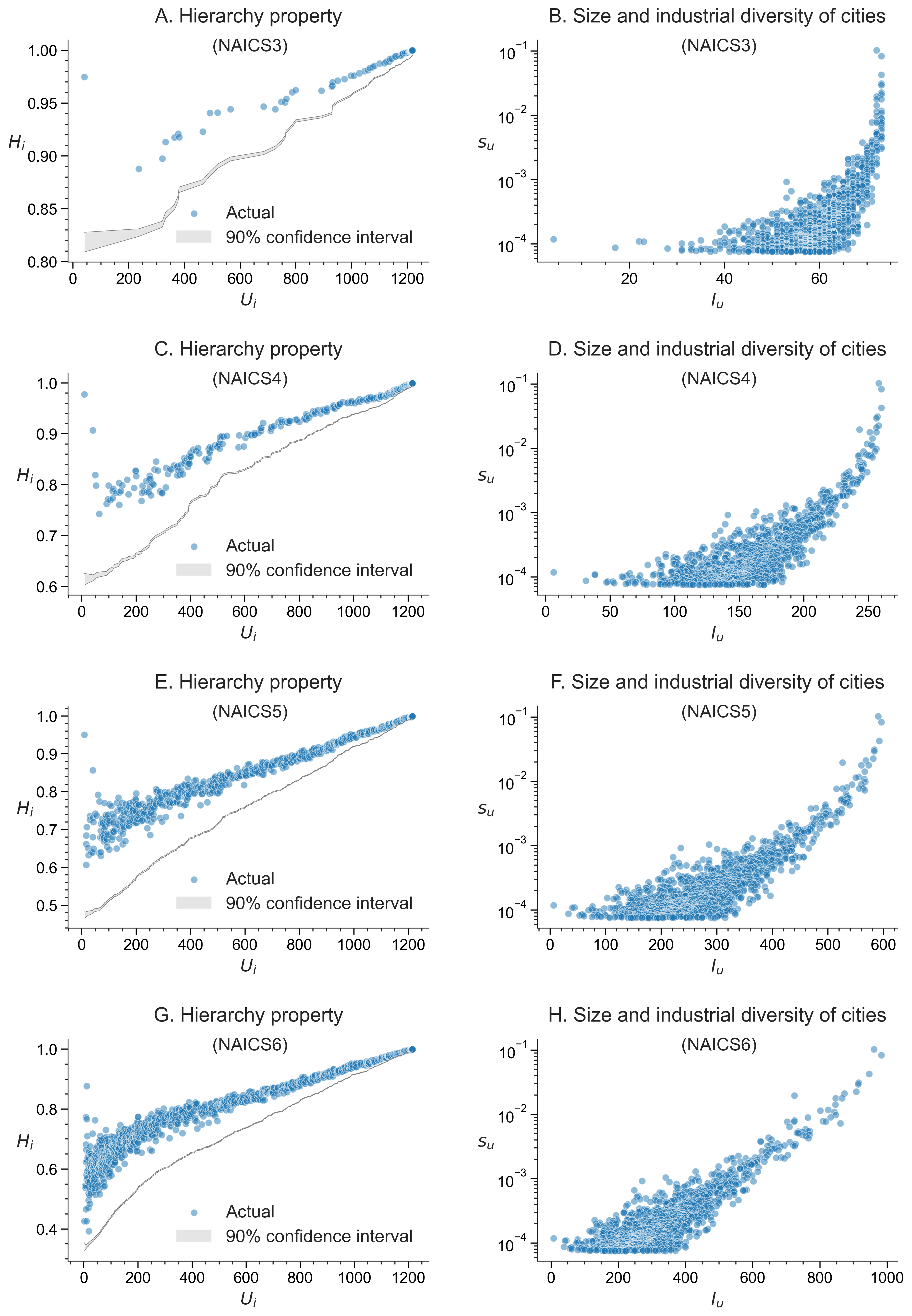}
    \caption{(A,\,C,\,E,\,G) Hierarchy shares $H_i$ plotted against the number $U_i$ of choice cities for the secondary and tertiary industries in the three-, four-, five- and six-digit NAICS, which include 72, 259, 597, and 999 industries, respectively, in 2000.
    The shaded area indicates the 90\% confidence interval of the counterfactual $\tilde{H}_i$ under the null hypothesis. 
    (B,\,D,\,F,\,H) The size $s_u$ and industrial diversity $I_u$ of cities $u$.}
    \label{fig:us_hp_2000}
\end{figure*}

Fig.\,\ref{fig:us_hp_2020}(A,\,C,\,E,\,G) and \ref{fig:us_hp_2000}(A,\,C,\,E,\,G) show the results of the HP test for secondary and tertiary industries in three-, four-, five-, and six-digit NAICS of the US in 2020 and 2000, respectively.
The size and industrial diversity of cities at each aggregation level are also shown in panels B, D, F, and H in each figure, indicating an obvious positive association between them.

\subsection{Empirical evidence for Japan\label{sec:si-jp}}
\subsubsection*{Cities} The number of cities is about one-third of that in the US, reflecting the similar ratio between the national population size of the two countries. 
The 431 cities identified in Japan in 2020 are shown in Fig.\,\ref{fig:jp}A.

\subsubsection*{Power laws} 
\begin{figure*}[h!]
\centering{}\includegraphics[width=17.1cm]{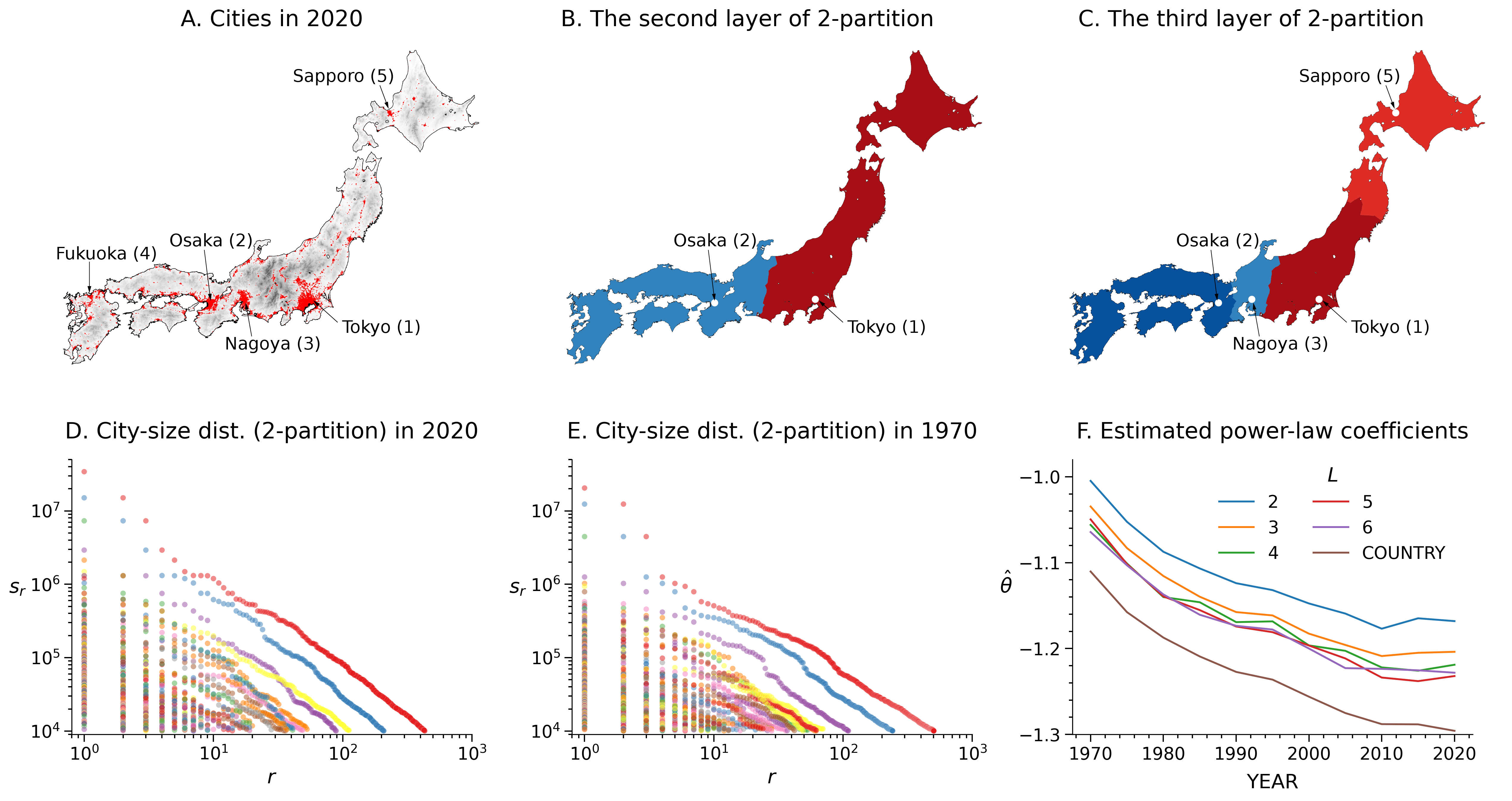}
\caption{(A) Red areas indicate all the 431 cities in Japan in 2020; darker gray corresponds to a larger population per 1 km-by-1 km grid outside cities. 
The 5 largest cities are indicated with their population rankings in parentheses. 
(B,\,C) The second and third layers of the 2-partition of cities in 2020.
(D,\,E) City-size distributions for the 2-partition cells of Japan in 2020 and 1970, respectively. (F) Estimated power-law coefficients, $\hat{\theta}$, under $L$-partition ($L=2,\ldots,6$)  together with that for the country in 1970--2020.
}
\label{fig:jp} 
\end{figure*}

Fig.\,\ref{fig:jp}B and C show the second and third layers of the 2-partition in 2020.
The city-size distributions for the 2-partition cells are shown in Fig.\,\ref{fig:jp}D and E for years 2020 and 1970, respectively. The CPL test rejects the null hypothesis at the 1\% level for all $L=2,\ldots,6$ in 2020 and 1970, except that it is significant at the 10\% level for $L=3$ in 1970.

Fig.\,\ref{fig:jp}F shows the persistent tendency of increasing concentration toward larger cities in Japan, so the values of the estimated power-law coefficients have steadily decreased throughout the period for the 2- to 6-partitions as well as for the country while preserving the CPL structure.
While a similar tendency is observed in the US (Fig.\,\ref{fig:us}B), it is stronger in Japan. 
One major reason is that transportation and communication costs plummeted substantially in Japan during this period, as high-speed railway and highway networks were built across the country almost from scratch.
The development of highways and high-speed railway networks in Japan was triggered by the Tokyo Olympics held in 1964. Between 1970 and 2020, the total highway (high-speed railway) length increased from 1,119 km (515 km) to 9,050 km (3,106 km), which is more than eight (six) times increase.
Moreover, the latter half of this period witnessed the introduction of the internet.
The theory of economic agglomeration predicts concentration toward fewer and larger cities when interregional transport costs decrease \citep[][]{Akamatsu-et-al-DP2021}.
In fact, the number of cities has also steadily decreased from 504 in 1970 to 431 in 2020.

\subsubsection*{Spatial-grouping property} 
\begin{figure*}
    \centering
    \includegraphics[width=.7\hsize]{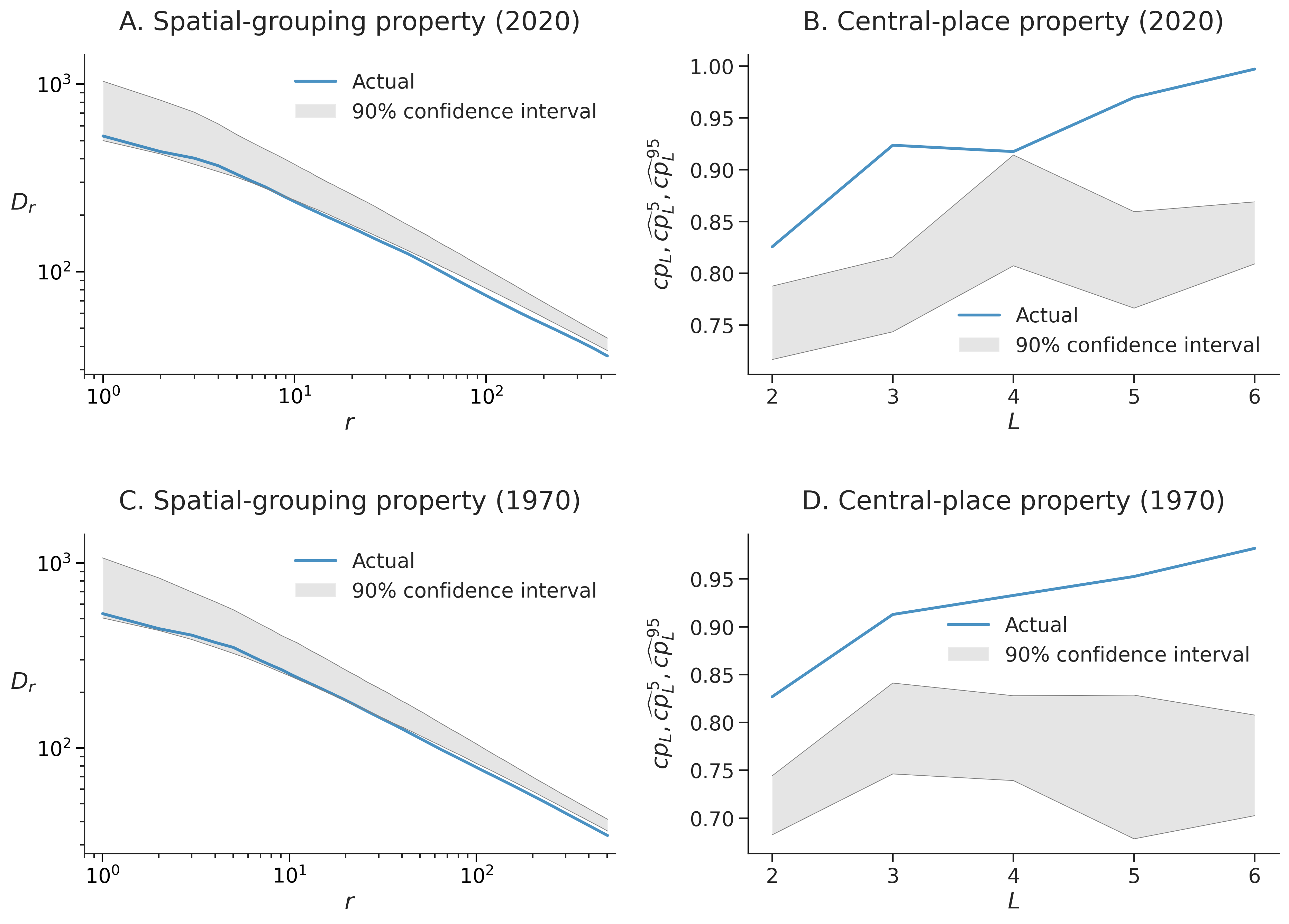}
    \caption{(A,\,C) The results of the spatial-grouping property tests for Japanese cities in 2020 and 1970, respectively. The shaded areas indicate the 90\% confidence intervals of the counterfactual $\tilde{D}_r$ under the null hypothesis. The null hypothesis is rejected for $r\geq 9$ and $r\geq 24$, respectively, at the 5\% level. (B,\,D) The results of the central-place property tests in 2020 and 1970, respectively. The null hypothesis is rejected for all the $L(=2,\ldots,6)$-partitions in both years.}
     \label{fig:cp_jp}
\end{figure*}

The results of the SGP test in 2020 and 1970 are shown in Fig.\,\ref{fig:cp_jp}A and C, respectively.
The null hypothesis that $D_r$ and $\tilde{D}_r$ belong to the same statistical population is rejected for $r\geq 9$ and $r\geq 24$, respectively, at the 5\% significance level. Similar results are obtained for the years 1975--2015. Specifically, for  1975, 1980, 1985, 1990, 1995, 2000, 2005, 2010, and 2015, the null hypothesis is rejected for all $r\geq 21, 7, 9, 7, 8, 8, 7, 6, 6$ and $9$, respectively.

\subsubsection*{Central-place property}
The CPP is persistent in Japanese cities. The null hypothesis is rejected at the 1\% level for all 2- to 6-partition cases in all years, 1970--2020. Fig.\,\ref{fig:cp_jp}B and D show the results in 2020 and 1970, respectively.

\subsubsection*{Hierarchy property}
\begin{figure*}
    \centering
    \includegraphics[width=.7\hsize]{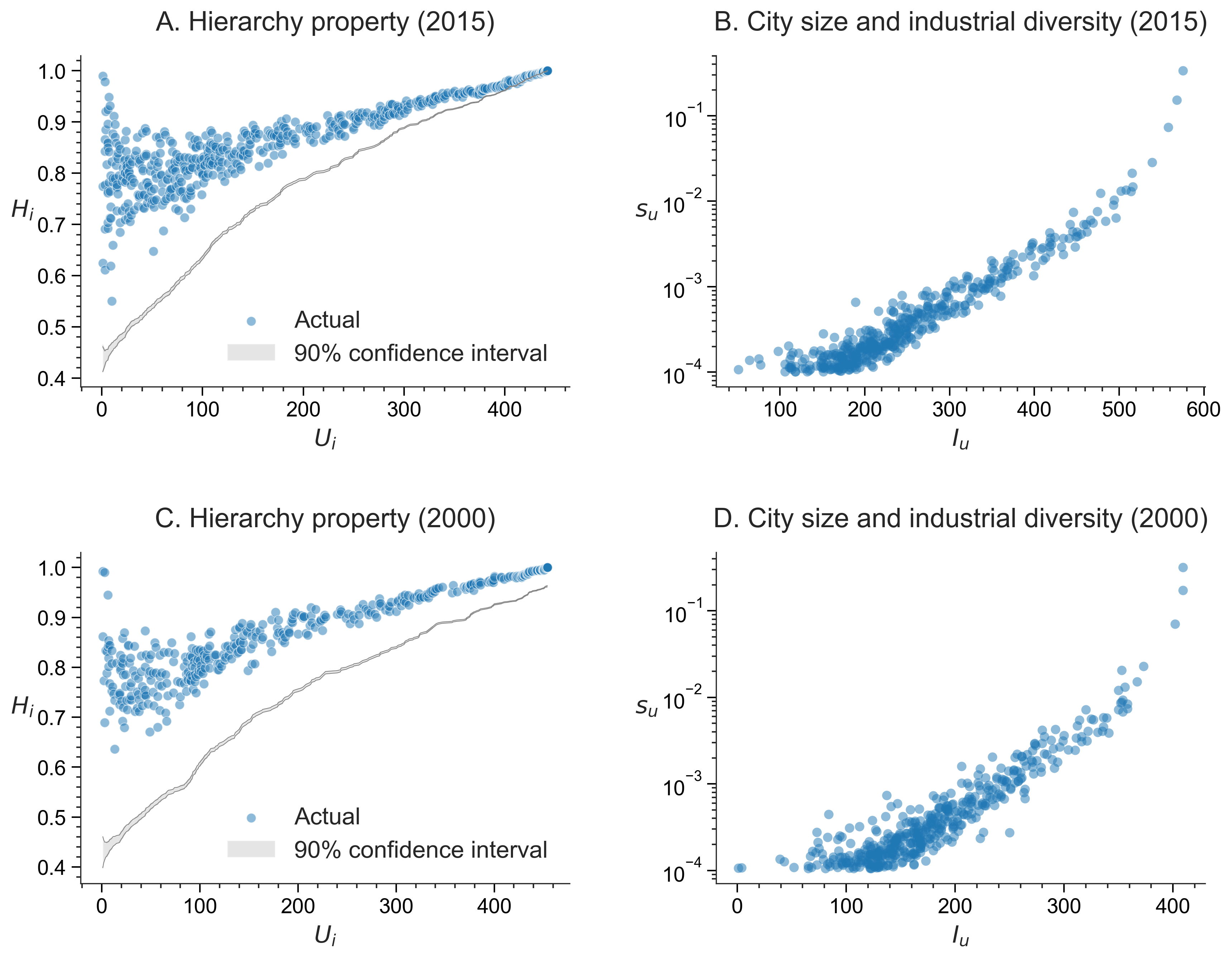}
    \caption{(A,\,C) $H_i$ against the number $U_i$ of choice cities for the three-digit JSIC secondary and tertiary industries together with the 90-percent confidence interval of $\tilde{H}_i$ in 2015 and 2000, respectively. The numbers of industries are 581 and 412, respectively. The null hypothesis ($H_i=\tilde{H}_i$) is rejected at the 5\% level for all industries in both years.
    (B,\,D) The relation between the size and industrial diversity of cities in 2015 and 2000, respectively. The Spearman's rank correlations are 0.933 and 0.898, respectively.}
     \label{fig:hp_jp}
\end{figure*}

Fig.\,\ref{fig:hp_jp}A  shows the results of the HP test in 2015. For all the 581 three-digit secondary and tertiary JSIC industries, the null hypothesis is rejected at the 1\% significance level in favor of the consistency with the HP. Similar results are shown in Fig.\,\ref{fig:hp_jp}C for the 412 industries in 2000. (The establishment locations can be identified at the 1km-by-1km grid level since 2000.)

The size and industrial diversity of cities in 2015 and 2000 are shown in Fig.\,\ref{fig:hp_jp}B and D, respectively. 
Their Spearman's rank correlations are 0.898 and 0.933, respectively, confirming a clear positive association between the two variables.

\subsubsection*{Volatility of city sizes and industrial location} 
\begin{figure*}
    \centering
    \includegraphics[width=.4\hsize]{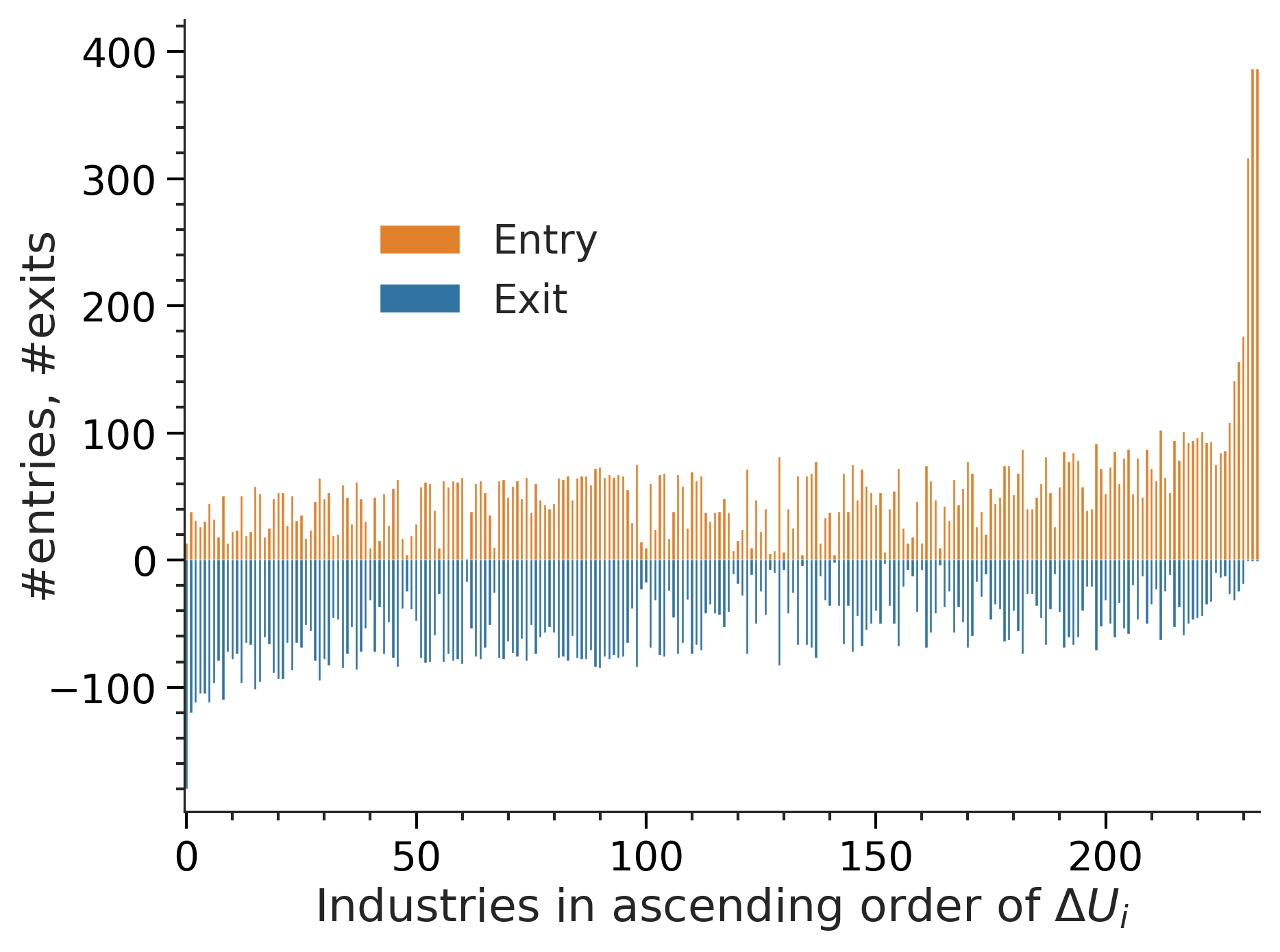}
    \caption{Churning of industries across cities. The numbers of choice cities in 2015 but not in 2000 (entry) and those in 2000 but not in 2015 (exit) of each of the 234 3-digit JSIC secondary and tertiary industries that existed in both years. Industries are ordered along the horizontal axis in the ascending order of the net change, $\Delta U_i$, in the number of choice cities between 2000 and 2015.}
    \label{fig:churning-jp}
\end{figure*}

Fig.~\ref{fig:churning-jp} shows the changes in the sets of choice cities of the 247 three-digit JSIC industries that existed in both 2000 and 2015. The numbers of entries to cities and exits from cities for industries between 2000 and 2015 are 55 and 54 on average, respectively.
Like the case of the US (Fig.\,\ref{fig:us}I), there are industries that exited from most of the cities during the 15 years. 
Table \ref{tb:industry-churning-jp} lists the 10 industries with the largest net decrease and those with the largest net increase in the number of choice cities between 2000 and 2015.
For example, ``Money Lenders'', ``Department Stores,'' and ``Musical Instrument Retailers'' are the three industries that had the least net entries between 2000 and 2015.
All of them are influenced more or less by the spread of the Internet.

\subsection{Supplementary simulation results\label{sec:si-sm}}
This section complements the discussion in the main text on the simulation results based on the theoretical model. 

\subsubsection*{Parameter values}
In all the simulations, we set the number of locations, $R=1\rm{,}024$, the mass of mobile workers, $H=1\rm{,}000$, and that of immobile workers, $L= 10\rm{,}000$.
The freeness-of-trade measure is given by $\phi_{i,rs} = \phi^{(\sigma_i -1)d_{rs}}\,\forall r,s\in\bm R$, where $d_{rs}$ represents the distance between locations $r$ and $s$ given by $\min\{|r-s|,R-|r-s|\}\times 2\pi/R$, and $\phi$ is the key parameter to be set.

In our model, the qualitative results largely depend on the level of transportation costs relative to the values of $H$ and $L$.
The theoretical results by \cite{Akamatsu-Takayama-Ikeda-JEDC2012,Akamatsu-et-al-DP2021} suggest that in a context of a model with a single industry subject to scale economies (i.e., $I=1$), a larger number of smaller cities are formed under larger transportation costs. 
Thus, the economy has a single city ($I$ cities) for a sufficiently small (large) value of $\phi$.
We set the $\phi=0.6$ so that the number of choice cities ranges from 1 to $R$, when a sufficiently large $I$ is chosen (e.g., $I=256$).%

\subsubsection*{Variation in scale economies and city size}
In our model, the variation in city size essentially accrues from that in scale economies among industries as Fig.~\ref{fig:size-diversity} indicates.
The number of cities is only 96 under $I=4$, whereas it is over 900 for $I=16,64,$ and 256.
At the same time, the diversity in city size substantially increases.
The ratio between the largest and the 95th percentile of the city sizes is 143, 96,672, 1,084,473, and 2,454,837 under $I=4,16,64,$ and $256$, respectively.

\subsubsection*{Spatial-grouping property}

\begin{figure*}
    \centering
    \includegraphics[width=.7\hsize]{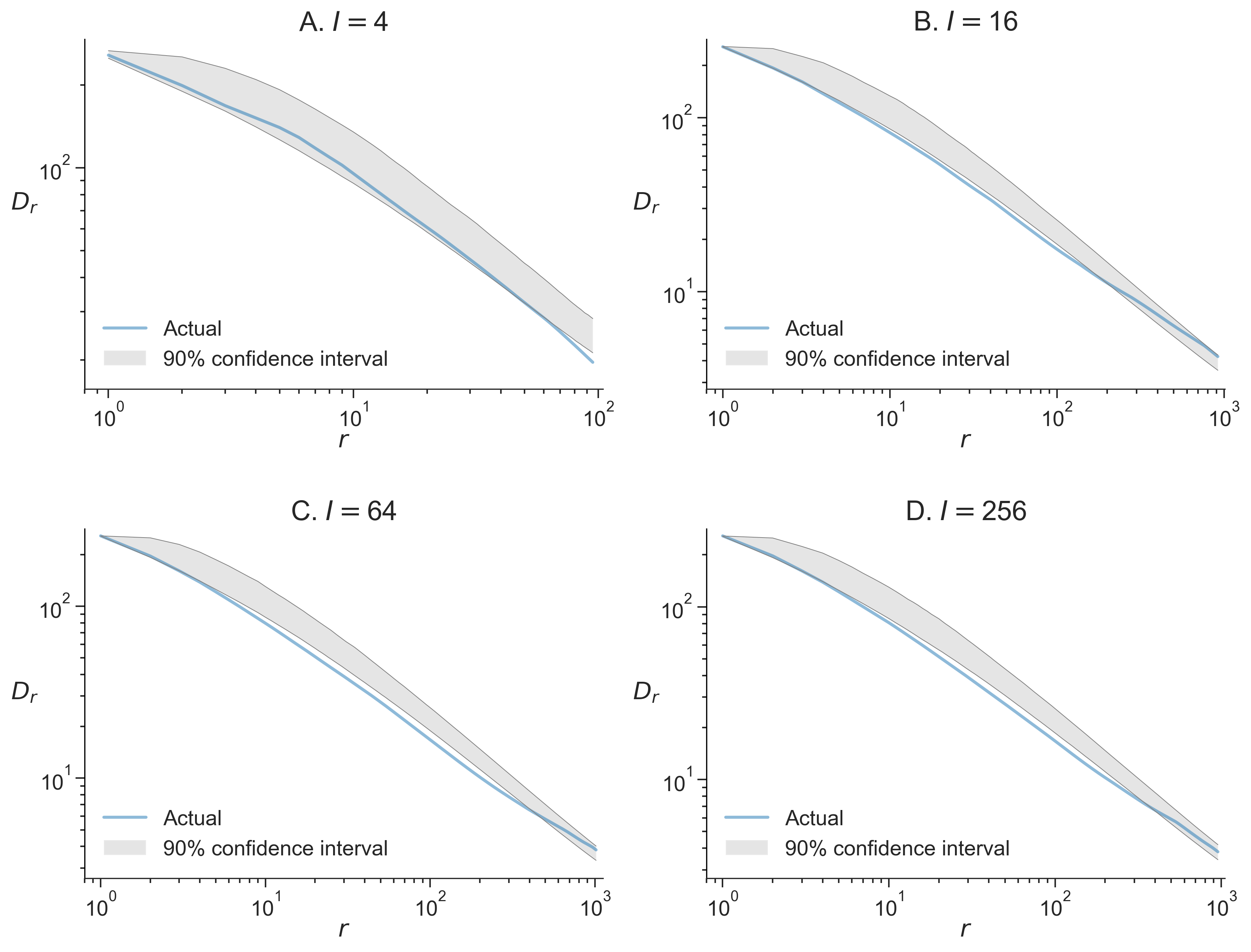}
    \caption{(A--D)
    $D_r$ together with the 90\% confidence interval of the counterfactual $\tilde{D}_r$ for the first equilibrium samples under $I=4,16,64,$ and 256, respectively ($R=1\rm{,}024$).}
    \label{fig:app-sgp-sm}
\end{figure*}

Fig.\,\ref{fig:app-sgp-sm}(A,\,B,\,C,\,D) show the SGP-test results (corresponding to Figs.~\ref{fig:us}E and \ref{fig:cplsm}C) for the first equilibrium sample (out of 1,000) for $I=4,16, 64$, and 256, respectively.
The confidence intervals are constructed from 1,000 samples of $\tilde{D}_r$ in each case.

As suggested by Fig.\,\ref{fig:size-diversity}, the city-size variation is substantially small under $I=4$, where the number of cities is 96 in this case, whereas it is more than 900 for $I=16,64,$ and 256.
As $I$ increases, the spatial pattern of cities is more consistent with the SGP so that they are placed more evenly over the spatial distribution of smaller cities.

The equilibrium under $I=4$ does not clearly exhibit SGP (Fig.~\ref{fig:app-sgp-sm}A), while those under $I=16$ and 64 exhibit SGP (Fig.~\ref{fig:app-sgp-sm}B,C) as in the case of $I=256$ (Fig.~\ref{fig:app-sgp-sm}D).
Equilibria under a larger $I$ is more consistent with the SGP. 
The SGP is significant
at the 5\% level for $r\in [4,190]$, $r\in [3,436]$, and $r\in [4,373]$ for $I=16$, 64, and 256, respectively.

Finally, we report the overall consistency of our equilibrium samples with the SGP.
For each of the 1,000 equilibrium samples under $I=4,16,64,128,$ and $256$, we compare $D_r$ (Eq.\ref{eq:SGP}) of the equilibrium and a random counterfactual value $\tilde{D}_r$ for each $r\leq 512$ ($=1\rm{,}024/2$) generated under the null hypothesis. 
Fig.\,\ref{fig:summary-sm}A shows, for each $(r,I)$, the share of the equilibrium samples consistent with the SGP (i.e., $D_r < \tilde{D}_r$) among all $1\rm{,}000$ equilibrium samples.
For the pair $(r,I)$ with the share exceeding $0.95$ (the dashed line), the null hypothesis is rejected in favor of the SGP at the 5\% level in the one-sided test.
The SGP is largely insignificant when the industrial diversity is small ($I=4$ and $16$). 
It is significant for $r < 300$ under a sufficiently large industrial diversity, $I\geq 64$.

\subsubsection*{Central-place property}
For each of the 1,000 equilibrium samples under given $L=2,\ldots,6$ and $I=4,16,64,128,$ and $256$, we compute $\textit{cp}_L$ of the equilibrium and a random counterfactual value $\tilde{\textit{cp}}_L$ generated under the random counterfactual partition of cities. 
Fig.\,\ref{fig:summary-sm}B shows, for each $(L,I)$, the share of the equilibrium samples that are consistent with the CPP (i.e., $\textit{cp}_L > \tilde{\textit{cp}}_L$) among the $1\rm{,}000$ equilibrium samples. 
For the pair $(L,I)$ with the share exceeding 0.95 (the dashed line), the null hypothesis is rejected in favor of the CPP at the 5\% level in the one-sided test.

\subsubsection*{Hierarchy property}
\begin{figure*}
    \centering
    \includegraphics[width=.4\hsize]{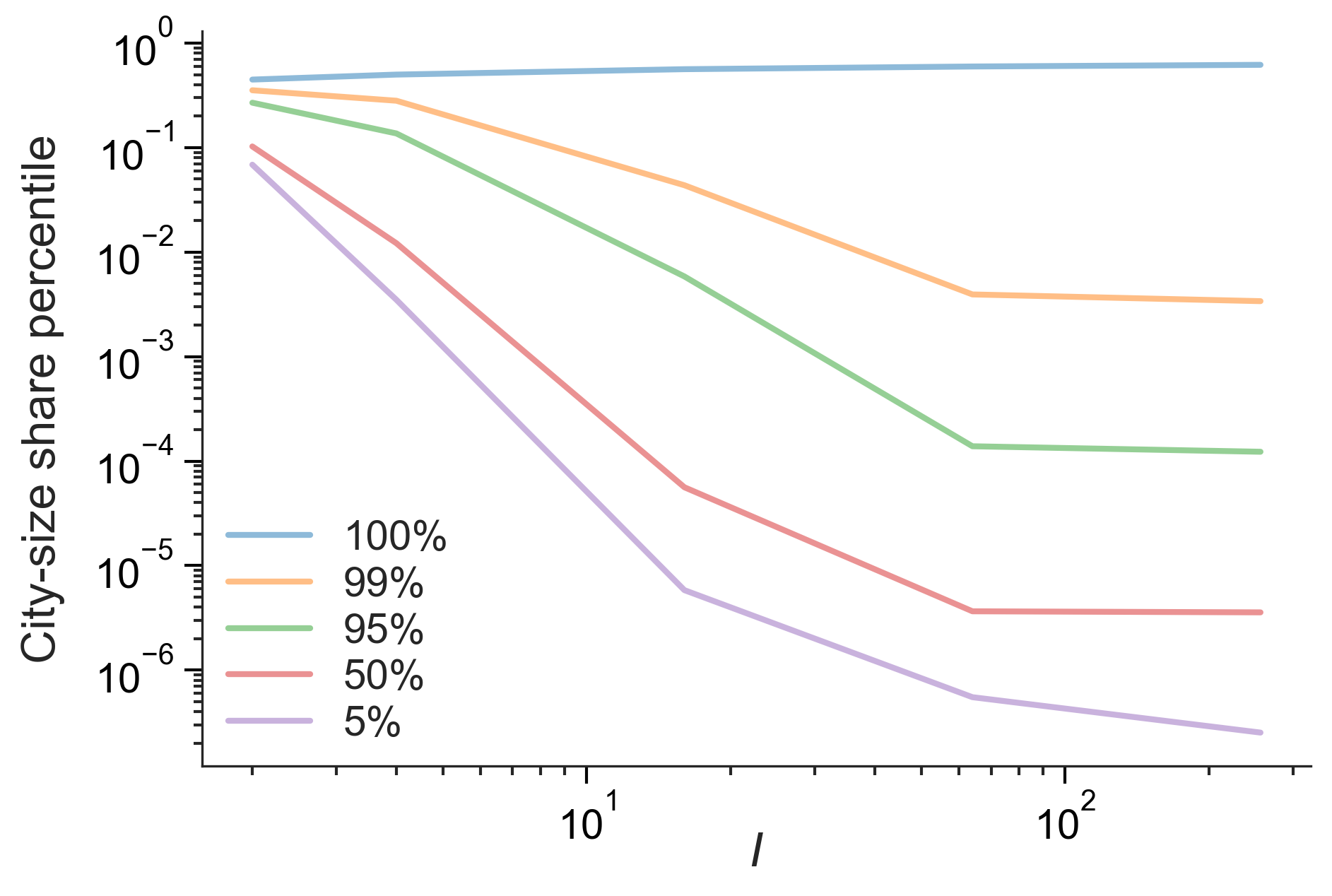}
    \caption{Percentiles of mobile-population shares of cities in 1,000 equilibrium samples under various $I$ with $R=1\rm{,}024$.}
    \label{fig:size-diversity}
\end{figure*}

\begin{figure*}
    \centering
    \includegraphics[width=.7\hsize]{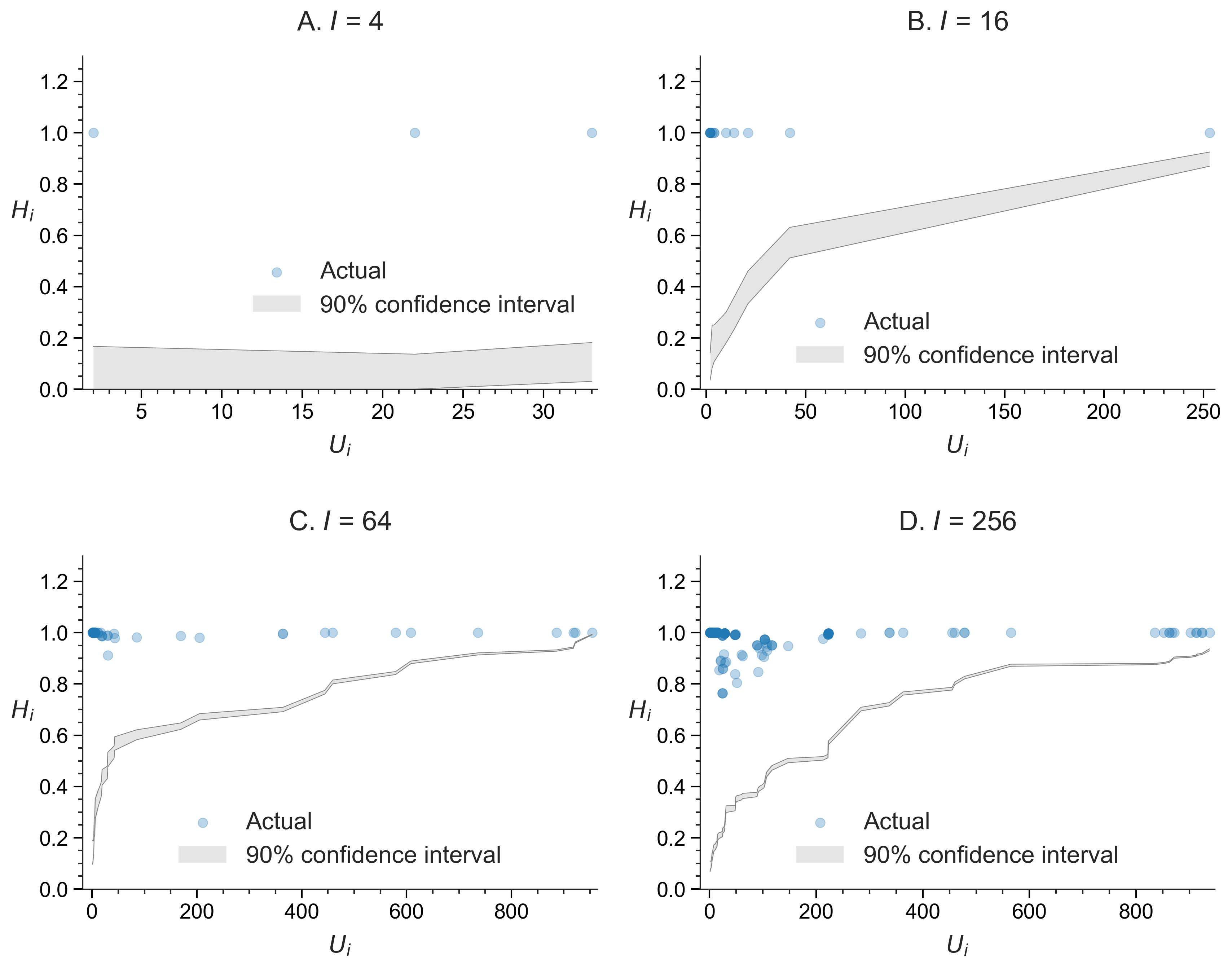}
    \caption{(A--D) Hierarchy shares $H_i$ against the number $U_i$ of choice cities in the first equilibrium sample for $I=4,16,64,$ and $256$, respectively ($R=1\rm{,}024)$, together with the 90\% confidence interval of the counterfactual $\tilde{H}_i$.
    The mean values of $H_i$ across all $i\in \bm I$ are 1.0, 1.0, 0.996, and 0.986, respectively. 
    $H_i$'s are significantly larger than $\tilde{H}_i$ at the 1\% level for all industries in all cases of $I=4, 16 , 64,$ and 256.}
    \label{fig:hp-sm}
\end{figure*}

\begin{figure*}
    \centering
    \includegraphics[width=.7\hsize]{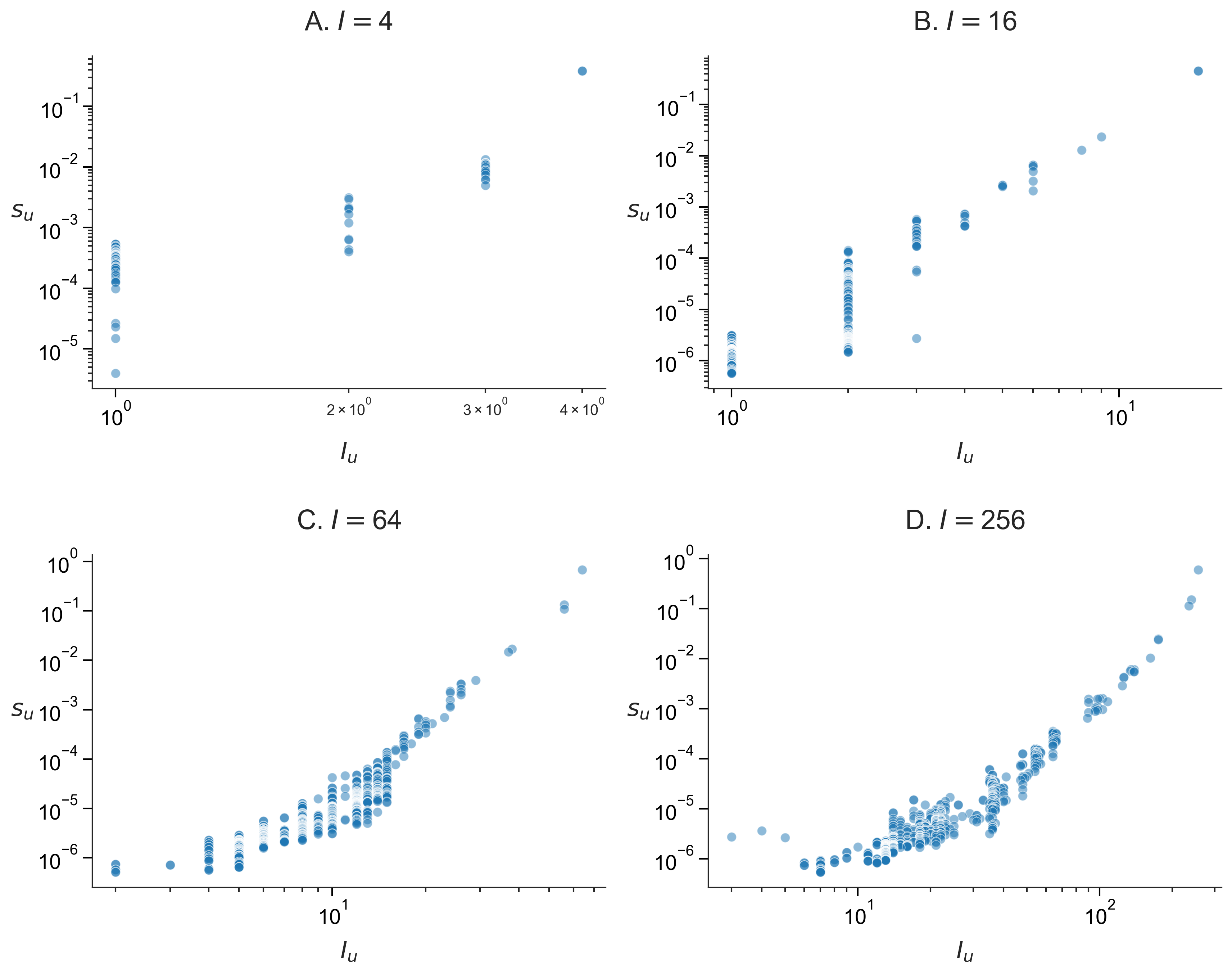}
    \caption{(A--D) Size $s_u$ of city $u$ against its industrial diversity $I_u$ in the first equilibrium sample for $I=4,16,64,$ and $256$, respectively ($R=1\rm{,}024$). 
    Spearman's rank correlations between $s_u$ and $I_u$ are 0.831, 0.754, 0.945, and 0.933, respectively.}
    \label{fig:div-sm}
\end{figure*}

\begin{figure*}
    \centering
    \includegraphics[width=\hsize]{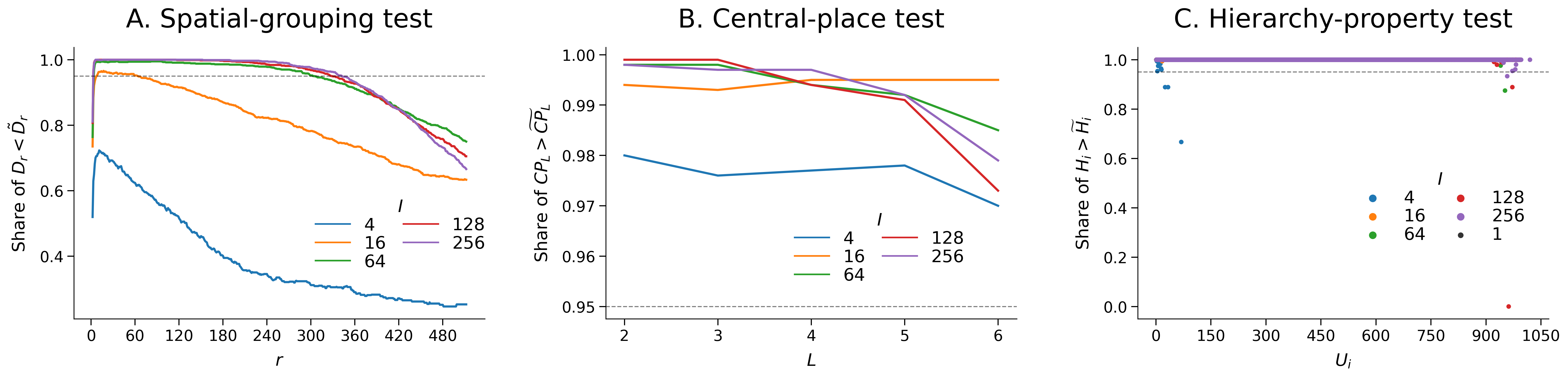}
    \caption{(A,\,B) For each of the 1,000 equilibrium samples, we compare a given test statistic $X$ of the equilibrium and one random counterfactual value $\tilde{X}$ generated under the null hypothesis of the test. 
    Panels A and B plot the shares of equilibrium samples under given values of $r$ and $L$ that are consistent with the alternative hypothesis of the test, $D_r < \tilde{D}_r$ and  $cp_L > \tilde{cp}_L$, for the SGP and CPP, respectively. In the cases in which the shares exceed 0.95, the one-sided test rejects the null hypothesis ($D_r=\tilde{D}_r$, $cp_L = \tilde{cp}_L$) at the 5\% level in favor of the SGP and CPP.
    (C) For the HP, we pool all industries $i$ with a given value of $U_i$ across all the 1,000 equilibrium samples under a given value of $I$. The panel plots the share of instances consistent with the HP, i.e., $H_i > \tilde{H}_i$ for each given value of $U_i$ in equilibrium samples. In the cases in which the shares exceed 0.95, the one-sided test rejects the null hypothesis ($H_i = \tilde{H}_i$) at the 5\% level in favor of the HP. 
    }
    \label{fig:summary-sm}
\end{figure*}

Figs.\,\ref{fig:hp-sm} (and Fig.\,\ref{fig:summary-sm}C) and \ref{fig:div-sm} suggest the mechanism underlying the positive correlation between industrial diversity and city size. 
On the one hand, the HP holds nearly perfectly irrespective of the industrial diversity $I$ of the economy (Fig.\,\ref{fig:hp-sm}, Fig.\,\ref{fig:summary-sm}C).
On the other hand, larger industrial diversity, $I$, allows a wider variation in city size, depending on the number of industries that happen to locate in a given city (Fig.~\ref{fig:size-diversity}).
Thus, the size of a city becomes a more precise indicator of its industrial diversity as $I$ increases.
This finding is reflected in the fact that the variation in city size, $s_u$, for a given industrial diversity, $I_u$, of city $u$ becomes smaller as $I$ increases (Fig.\,\ref{fig:div-sm}). 

To evaluate the overall consistency of our equilibrium samples with the HP, we compare hierarchy share $H_i$ (Eq.\,\ref{eq:h}) and a random counterfactual value $\tilde{H}_i$ for each $i\in \bm I$ generated under the null hypothesis. 
As $\{\sigma_i\}$ are randomly drawn for each equilibrium sample, the number of cities each industry with index $i$ locates, $U_i$, can differ across 1,000 equilibrium samples for each $i$. 
We use the number of industry-choice cities to construct the set of sample industries for testing the HP. 
For each equilibrium samples $k\in\{1,\ldots,1\rm{,}000\}$, let $i(k)$ indicate industry $i$ in equilibrium sample $k$.
For each number $u = 1,2,\hdots,U$ of industry-choice cities, we pool all industries $i(k)$ that satisfies $U_{i(k)} = u$ across all the equilibrium samples, $k=1,\ldots,1\rm{,}000$, under each  $I=4,16,64,128,$ and $256$.
Specifically, for each $u$, we consider a set $\bm I_u \equiv \cup_{k = 1}^{1000} \{i(k) \in \bm I \mid u = U_{i(k)}\}$. 
Then, we compute 
the share of instances consistent with the HP, i.e., $H_i > \tilde{H}_i$ for each $u$, that is, $|\{i(k)\in \bm I_u \mid H_{i(k)} > \tilde{H}_{i(k)}\}|/|\bm I_u|$. 
Fig.\,\ref{fig:summary-sm}C plots these shares against the realized values of the number of choice cities, $U_i = u$ in equilibrium samples.
For a pair $(U_i,I)$ with the share exceeding 0.95 (the dashed line), the null hypothesis is rejected in favor of the HP at the 5\% level in the one-sided test.
The HP is significant except for a few highly localized and ubiquitous industries.

\begin{figure*}
    \centering
    \includegraphics[width=.6\hsize]{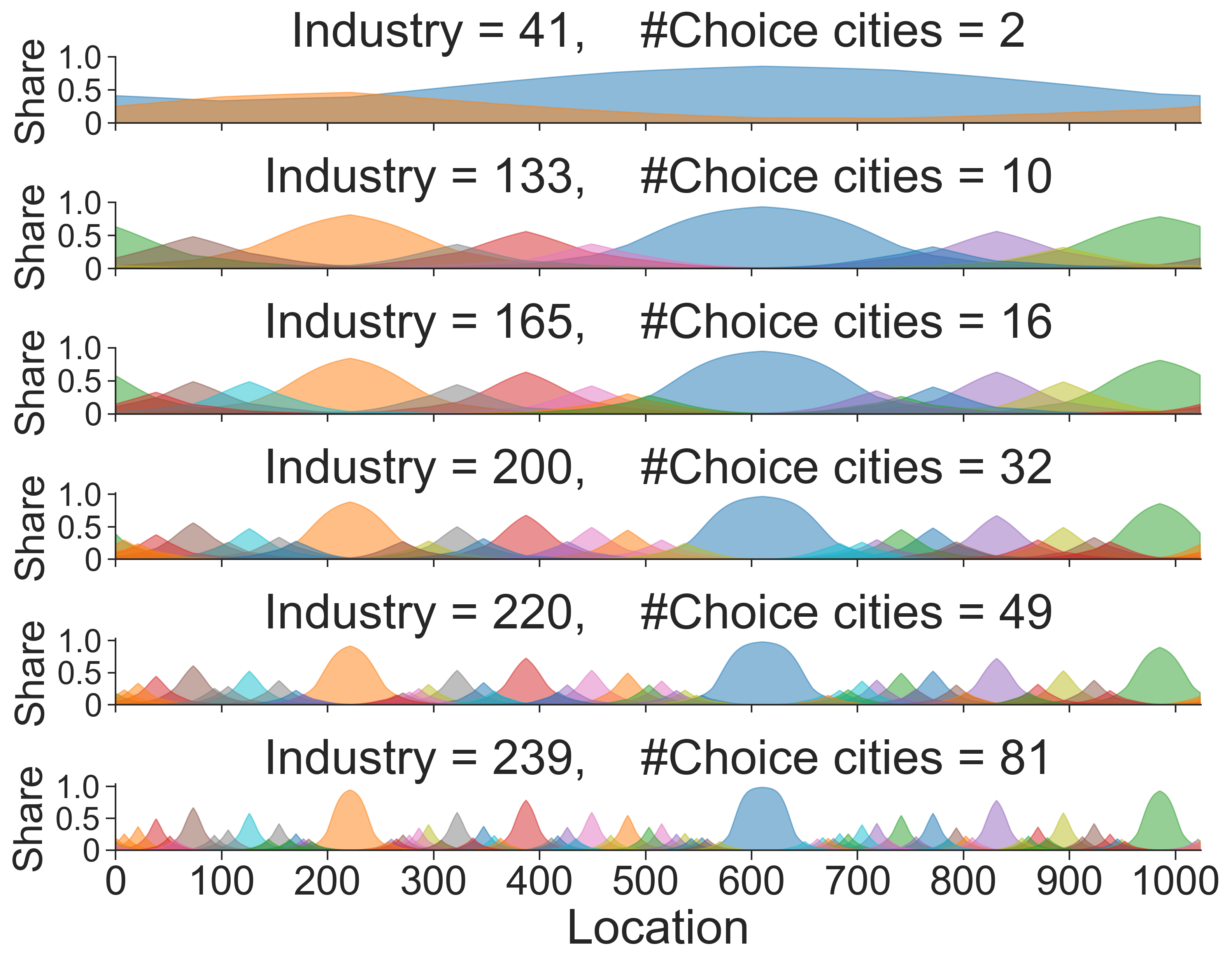}%
    \caption{Choice cities' market areas for selected industries. Each panel depicts, for a designated industry, the share of each choice city in the market at each location, where the same color indicates the same choice city. The market areas are shown only for choice cities with the largest market share in at least one location.}
    \label{fig:market-area}
\end{figure*}

Finally, we demonstrate that the spatial coordination of industries implied by the HP results in the SGP.
For the first sample of the equilibrium under $I=256$ discussed in Fig.\ref{fig:sm}, Fig.\,\ref{fig:market-area} shows the market shares of choice cities in each location.
The figure indicates that the spacing of central places in each layer corresponds to the size of the market area of the layer-specific industry, revealing the mechanism behind the SGP.
The market share of $i$-choice city $r$ in location $s$ is expressed by
\begin{equation}
    M_{i,rs} = \frac{T_{i,rs}}{\sum_{k\in\bm R} T_{i,ks}}, 
\end{equation}
where $T_{i,rs}$ is the value of industry-$i$ exports from $r$ to $s$, and is given by 
\begin{equation}
    T_{i,rs}= p_{i}\tau_{rs}Q_{i,rs}n_{ir} = \frac{\phi_{i,rs}h_{ir}}{\Delta_{is} (\bm h_i)} (h_s+l) 
\end{equation}
from Eqs.\, \ref{eq:price-index}, \ref{eq:demand}, \ref{eq:price}, \ref{eq:total-demand}, \ref{eq:price-index-2}, and the relation, $n_{ir}=h_{ir}$.

\begin{figure*}
    \centering
    \includegraphics[width=.8\hsize]{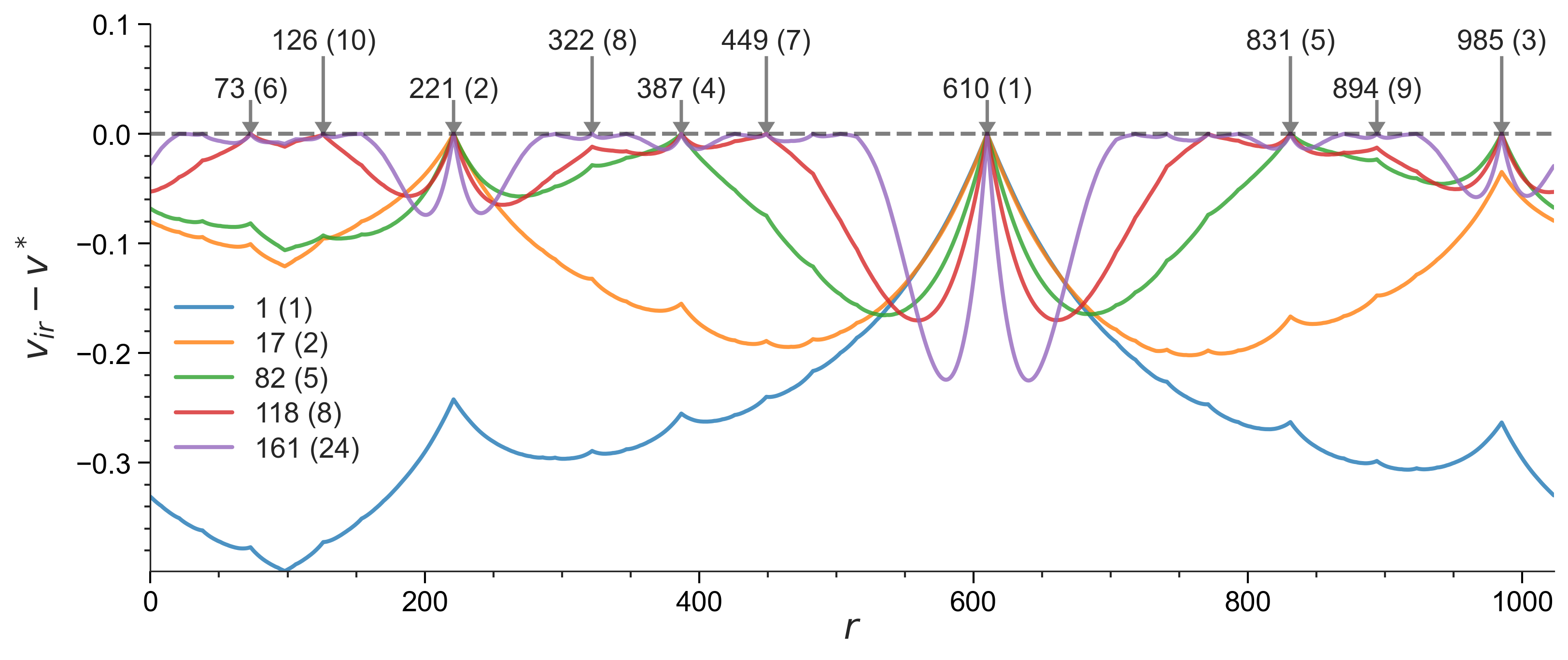}
    \caption{The relative utility level $v_{ir} - v^*$ of mobile workers employed in the selected industries, $i=1, 17, 82, 118,$ and 161 in each location $r$ in the first equilibrium sample under $(I,R)=(256,1\rm{,}024)$.
    The numbers in parentheses in the legend are the numbers of choice cities of these industries.
    The locations of the largest 10 cities are indicated by arrows, where the numbers in parentheses are the size-ranking of cities.
    }  
    \label{fig:utility-sm}
\end{figure*}

The spatial coordination among industries can also be visualized by the  utility levels of mobile workers employed in each industry.
Fig.\,\ref{fig:utility-sm} shows the case of the selected industries in the first equilibrium sample under $(I,R)=(256,1\rm{,}024)$. 
If $v_{ir} = v^*$, then city $r$ is a $i$-choice city, whereas location $r$ offers at most negative profit if $v_{ir} < v^*$. 
Notice that larger cities are choice cities of a larger number of industries. 
A kink of the utility curve at a city location indicates the agglomeration effect, so that the deviation from the city location causes the utility level (as well as the profit level) to deteriorate immediately.

\begin{table*}[h!]
\caption{Industries with the largest changes in the number of locations in the US}
\label{tb:industry-churning}
\begin{tabular*}{\hsize}{rllrrrr}
\toprule[1.5pt]
Rank&NAICS code&Description&$U_{i,2020}$&$\Delta U_{i}$&$\Delta_+ U_i$&$\Delta_- U_i$.\\
\midrule[0.6pt]\\[-2pt]
\multicolumn{7}{c}{A. 10 industries with the largest net decrease in the number of choice cities}\\[5pt]
1&5323&General Rental Centers&31&$-$928&4&$-$932\\
2&4512&Book Stores and News Dealers&221&$-$878&25&$-$903\\
3&6221&General Medical and Surgical Hospitals&95&$-$846&6&$-$852\\
4&4531&Florists&361&$-$823&58&$-$881\\
5&4542&Vending Machine Operators&80&$-$745&10&$-$755\\
6&3371&Household and Institutional Furniture and Kitchen Cabinet Manufacturing&266&$-$725&52&$-$777\\
7&5615&Travel Arrangement and Reservation Services&442&$-$719&87&$-$806\\
8&3273&Cement and Concrete Product Manufacturing&323&$-$681&78&$-$759\\
9&3152&Cut and Sew Apparel Manufacturing&31&$-$676&1&$-$677\\
10&5111&Newspaper, Periodical, Book, and Directory Publishers&418&$-$666&94&$-$760\\[8pt]
\midrule[0.6pt]\\[-3pt]
\multicolumn{7}{c}{B. 10 industries with the largest net increase in the number of choice cities}\\[5pt]
223&6212&Offices of Dentists&1279&65&441&$-$376\\
224&4471&Gasoline Stations&1288&72&444&$-$372\\
225&5221&Depository Credit Intermediation&1288&73&443&$-$370\\
226&5242&Agencies, Brokerages, and Other Insurance Related Activities&1286&74&444&$-$370\\
227&6213&Offices of Other Health Practitioners&1284&76&442&$-$366\\
228&5311&Lessors of Real Estate&1286&76&446&$-$370\\
229&5412&Accounting, Tax Preparation, Bookkeeping, and Payroll Services&1285&76&444&$-$368\\
230&8111&Automotive Repair and Maintenance&1291&77&448&$-$371\\
231&5617&Services to Buildings and Dwellings&1296&80&454&$-$374\\
232&8131&Religious Organizations&1292&80&448&$-$368\\[2pt]
\midrule[1.0pt]
\end{tabular*}
\setlength{\baselineskip}{9pt}
\addtabletext{(A,\, B) The list of ten four-digit NAICS industries that experienced
the largest net decrease and increase, respectively, in the number of choice cities between 2000 and 2020. ``Rank'' indicates the industry ranking in terms of the net change, $\Delta U_i$, in the number of choice cities between 2000 and 2020, in ascending order. 
$U_{i,t}$ is the number of choice cities for industry $i$ in year $t$. $\Delta_+ U_i\equiv |\bm U_{i,2020}\backslash \bm U_{i,2000}|$ and $\Delta_- U_i\equiv |\bm U_{i,2000}\backslash \bm U_{i,2020}|$.}
\end{table*}

\begin{table*}[h!]
\caption{Industries with the largest changes in the number of locations in Japan}
\label{tb:industry-churning-jp}
\begin{tabular*}{\hsize}{rllrrrr}
\toprule[1.5pt]
Rank&JSIC code&Description&$U_{i,2015}$&$\Delta U_{i}$&$\Delta_+ U_i$&$\Delta_- U_i$.\\
\midrule[0.6pt]\\[-2pt]
\multicolumn{7}{c}{A. 10 industries with the largest net decrease in the number of choice cities}\\[5pt]
1&641&Money Lending Business&139&$-$167&13&$-$180\\
2&561&Department Stores and General Merchandise Supermarkets&235&$-$82&38&$-$120\\
3&60C&Musical Instrument Stores&277&$-$81&31&$-$112\\
4&784&Public Bathhouses&205&$-$79&26&$-$105\\
5&301&Communication Equipment and Related Products&110&$-$75&30&$-$105\\
6&60B&Toy and Amusement Goods Stores&348&$-$68&44&$-$112\\
7&651&Financial Products Transaction Dealers&235&$-$68&32&$-$97\\
8&117&Underwear&88&$-$61&18&$-$79\\
9&121&Sawing, Planing and Wood Products&297&$-$60&50&$-$110\\
10&672&Non-life Insurance Institutions&224&$-$59&13&$-$72\\[8pt]
\midrule[0.6pt]\\[-3pt]
\multicolumn{7}{c}{B. 10 industries with the largest net increase in the number of choice cities}\\[5pt]
225&55A&Agents and Brokers&150&60&93&$-$33\\
226&795&Crematories and Graveyard Custodians&99&65&75&$-$10\\
227&805&Public Gardens and Amusement Parks&120&70&84&$-$14\\
228&661&Financial Auxiliaries&141&73&86&$-$13\\
229&118&Japanese Style Apparel, Other Textile Apparel, and Accessories&190&81&108&$-$27\\
230&694&Retail Estate Managers&378&109&141&$-$32\\
231&80A&Sports Facilities&190&131&156&$-$32\\
232&911&Employment Services&354&157&176&$-$19\\
233&416&Services Incidental to Video Picture Information, Sound Information, &&&\\
&&\hspace{35pt}Character Information Production, and Distribution&338&315&386&$-$1\\
234&80B&Sports, Amusement and Recreation Facilities&409&385&386&$-$1\\[2pt]
\midrule[1.0pt]
\end{tabular*}
\setlength{\baselineskip}{9pt}
\addtabletext{(A,\, B) The list of ten four-digit NAICS industries that experienced
the largest net decrease and increase, respectively, in the number of choice cities between 2000 and 2020. ``Rank'' indicates the industry ranking in terms of the net change, $\Delta U_i$, in the number of choice cities between 2000 and 2020, in ascending order. 
$U_{i,t}$ is the number of choice cities for industry $i$ in year $t$. $\Delta_+ U_i\equiv |\bm U_{i,2020}\backslash \bm U_{i,2000}|$ and $\Delta_- U_i\equiv |\bm U_{i,2000}\backslash \bm U_{i,2020}|$.}
\end{table*}

